\pdfoutput=1
\pdfoutput=1
\documentclass[aps,amsmath,amssymb,preprintnumbers,preprint,nofootinbib,a4paper,11pt]{article}
\usepackage{epsfig}
\usepackage{jheppub,undertilde}
\usepackage{amsthm}
\usepackage{graphicx}
\usepackage{color}
\usepackage{graphicx,textcomp,float,slashed}
\usepackage{arydshln}
\usepackage{empheq}
\usepackage{slashed}
\usepackage{feynmp}
\usepackage{hhline}

\newcommand{\ds}{\displaystyle×}

\newcommand{\be}{\begin{equation}}
\newcommand{\ee}{\end{equation}}

\def\bsp#1\esp{\begin{split}#1\end{split}}
\def\bpm{\begin{pmatrix}}
\def\epm{\end{pmatrix}}

\newcommand{\bea}{\begin{eqnarray}}  
\newcommand{\eea}{\end{eqnarray}}  
 \def\bsp#1\esp{\begin{split}#1\end{split}}
\preprint{\begin{flushright} KCL-PH-TH/2018-35

IFT-UAM/CSIC-18-77

IPPP/18/56\end{flushright}}

\title{Signs of heavy Higgs bosons at CLIC:\\
An $e^+ e^-$ road to the Electroweak Phase Transition}

\author[1,2]{J.~M.~No}
\author[3]{and M.~Spannowsky}

\affiliation[1]{Department of Physics, King's College London, Strand, WC2R 2LS London, UK}
\affiliation[2]{Departamento de Fisica Teorica and Instituto de Fisica Teorica, IFT-UAM/CSIC,
Cantoblanco, 28049, Madrid, Spain}
\affiliation[3]{Institute of Particle Physics Phenomenology, Physics Department, Durham University,
Durham DH1 3LE, UK}
\emailAdd{Josemiguel.no@uam.es}
\emailAdd{Michael.Spannowsky@durham.ac.uk}

\abstract{We analyse the sensitivity of the proposed Compact Linear Collider (CLIC) to the existence of 
beyond the Standard Model (SM) Higgs bosons through their decays into pairs of massive gauge bosons 
$H \to VV$ and SM-like Higgses $H \to hh$, considering 
CLIC centre of mass energies 
$\sqrt{s} = 1.4$ TeV and $3$ TeV. 
We find that resonant di-Higgs searches at CLIC would allow for up to two orders of magnitude improvement w.r.t.~the sensitivity achievable by 
HL-LHC in the mass range $m_H \in [250\,\mathrm{GeV},\, 1 \,\mathrm{TeV}]$.
Focusing then on a real singlet extension of the SM, we explore the prospects of heavy Higgs searches at CLIC 
for probing the regions of parameter space yielding a strongly first order electroweak phase transition 
that could generate the observed matter-antimatter asymmetry of the Universe. 
Our study illustrates the complementarity between CLIC and other possible future colliders like FCC{\sl-ee} in probing singlet extensions of the SM, and 
shows that high-energy $e^+ e^-$ colliders provide a powerful means to 
unravel the nature of electroweak symmetry breaking in the early Universe.}

\notoc

\begin{document}

\maketitle

\newpage

\tableofcontents

\section{Introduction}

A key goal of the present and future collider physics programme is to reveal the structure of the (scalar) sector responsible 
for electroweak symmetry breaking (EWSB) in Nature. While ongoing ATLAS and CMS analyses at the Large Hadron Collider (LHC) show that the properties of the 
discovered Higgs particle are close to those expected for the Standard Model (SM) 
Higgs boson $h$~\cite{Khachatryan:2016vau,ATLAS:2017ovn,CMS:2018lkl}, it still needs to be determined 
whether the scalar sector is realised in its most minimal form, i.e. consisting of one $SU(2)_L$ doublet, or has a richer structure, containing additional states. 
Non-minimal scalar sectors are very well-motivated, arising 
naturally in the context of weakly coupled completions of the SM that address the hierarchy 
problem.
At the same time, extensions of the SM scalar sector could provide the means 
to address a key open question at the interface of particle physics and cosmology, namely the 
generation of the cosmic matter-antimatter asymmetry, via electroweak (EW)
baryogenesis~\cite{Morrissey:2012db}.

Among the proposed future collider experiments, the Compact Linear Collider (CLIC) would be a 
multi-TeV $e^+ e^-$ collider~\cite{Aicheler:2012bya,CLIC:2016zwp}, combining the high-energy reach with the clean collision 
environment of an electron-positron machine. CLIC would operate in three energy stages, corresponding to centre of mass (c.o.m.) energies 
$\sqrt{s} = 380$ GeV, $1.4$ TeV, $3$ TeV, providing an ideal setup to study the properties of the Higgs sector.  
In this respect, very sensitive direct probes of the existence of new, heavier Higgs bosons, possible 
with $\sqrt{s} = $ $1.4$ TeV and $3$ TeV c.o.m.~energy configurations,
are highly complementary to precise measurements of the properties of the 125 GeV Higgs boson, 
and may yield the dominant probe of a non-standard Higgs sector.

In this work we analyse the reach of CLIC in searching for heavy Higgs bosons which decay to a pair of massive gauge bosons $VV = W^+W^-, Z Z$ or a pair of 
125 GeV Higgs bosons.
This allows to assess the direct sensitivity of CLIC to non-minimal Higgs sectors, and to compare it with that of the HL-LHC, providing at the same time 
a benchmark for sensitivity comparison with other possible future high-energy collider facilities like FCC({\sl-ee} and {\sl-hh}).  
In addition, we assess the capability of CLIC heavy Higgs searches in probing the nature of the EW phase transition in
the context of a general real singlet scalar extension of the SM~\cite{Profumo:2007wc,Barger:2007im,Espinosa:2011ax}.  
This scenario can capture the phenomenology of the Higgs sector in 
more complete theories beyond the SM such as the NMSSM (see~\cite{Ellwanger:2009dp} and references therein) 
or Twin Higgs theories~\cite{Chacko:2005pe}.
At the same time, the singlet scalar extension of the SM constitutes a paradigm 
for achieving a strongly first order EW phase transition that could generate the observed matter-antimatter asymmetry of the Universe.

\vspace{1mm}
The paper is organised as follows: In Section~\ref{subsec:CLIC} we discuss the main aspects of Higgs production at CLIC, as well 
as the various computational tools we use for our analysis. In Section~\ref{subsec:VV} we assess the CLIC sensitivity in direct searches 
of heavy scalars decaying into EW gauge boson pairs. In Section~\ref{subsec:hh} we focus instead on heavy scalar decays into 
a pair of 125 GeV Higgses. In Section~\ref{sec:xSM} we 
discuss the implications of these results for a singlet scalar extension of the SM, and the possibility of 
exploring the nature of the EW phase transition in this scenario via direct scalar searches at CLIC. 
Finally we conclude in Section~\ref{conclusionsNS}. 

\begin{figure}[h]

\begin{center}
\includegraphics[width=0.85\textwidth]{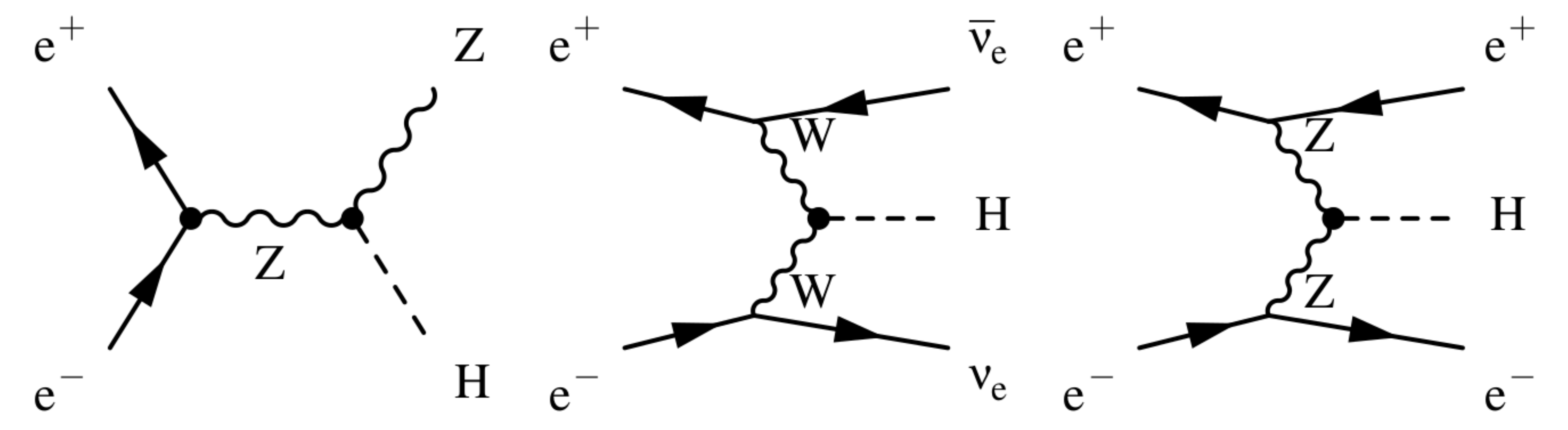}

\caption{\small Feynman diagrams for the three dominant Higgs boson production modes: 
$e^+ e^- \to H Z$ (left), $e^+ e^- \to H \nu \nu$ (middle) and $e^+ e^- \to H e^+ e^-$ (right).}
\label{HProd_CLIC_Feynman}
\end{center}

\vspace{-4mm}

\end{figure} 

\section{Heavy Higgs boson production at the Compact Linear Collider}
\label{subsec:CLIC}

\begin{figure}[h]

\begin{center}
\includegraphics[width=0.98\textwidth]{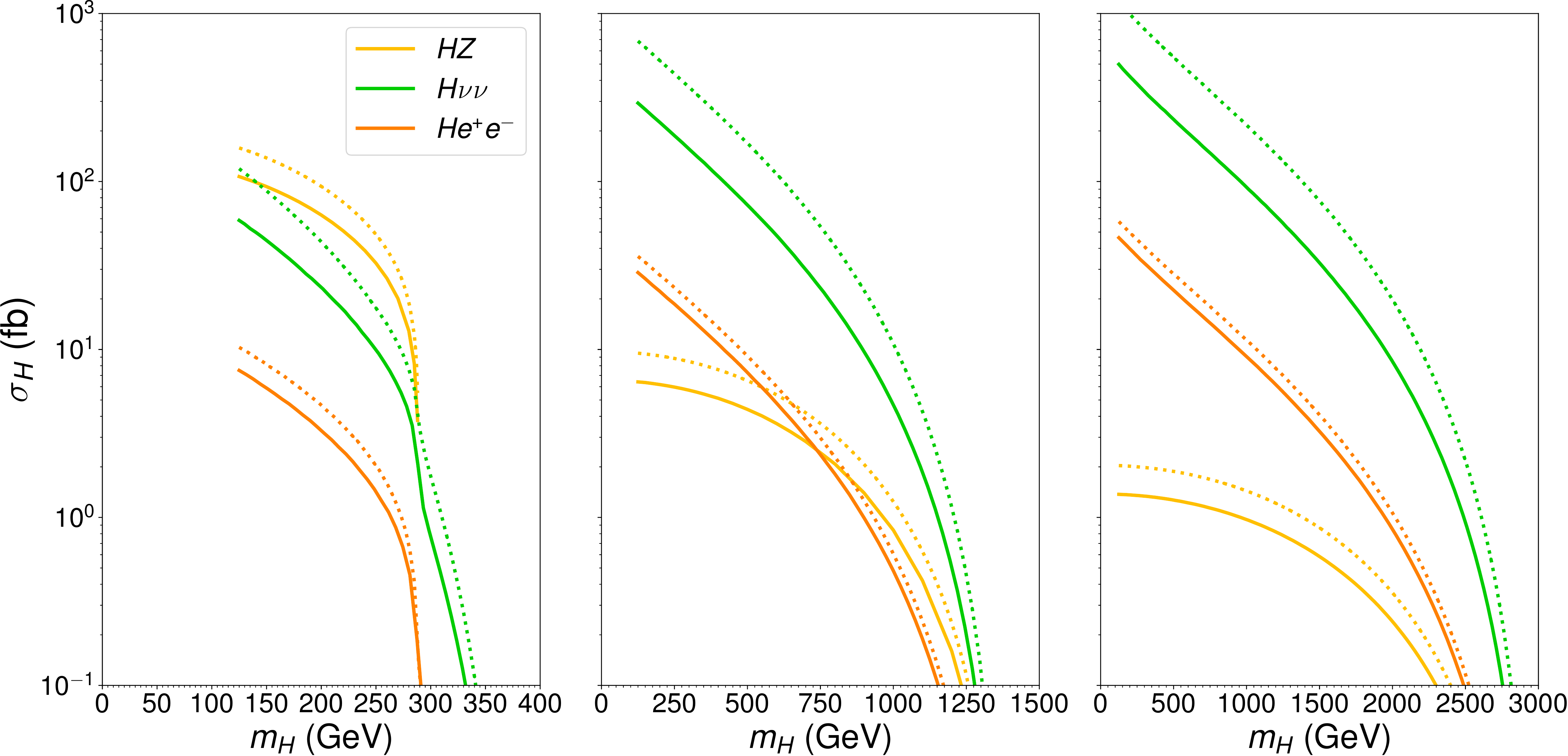}

\caption{\small Higgs production cross sections $\sigma_H$ (in fb), assuming SM-like properties for $H$, as a function of $m_H$, 
for $\sqrt{s} = 380$ GeV (left), $\sqrt{s} = 1400$ GeV (middle) and $\sqrt{s} = 3000$ GeV (right), 
for unpolarized beams (solid) and $P_{e^-},\,P_{e^+} = -80\%, \, + 30\%$ (dashed).}
\label{HProd_CLIC}
\end{center}

\vspace{-5mm}

\end{figure}

The three dominant processes contributing to Higgs boson production at a high-energy electron-positron collider are 
$e^+ e^- \to H\, Z$, $e^+ e^- \to H \nu \nu$ and $e^+ e^- \to H e^+ e^-$ (see e.g.~Figure~\ref{HProd_CLIC_Feynman}). 
Assuming a heavy scalar $H$ with SM-like properties, we compute the production cross 
section\footnote{For $e^+ e^- \to H e^+ e^-$, the outgoing electrons are required to satisfy $|\eta| < 5$, $P_T > 5$ GeV.} as a 
function of the scalar mass $m_H$ for each of the three processes and for $\sqrt{s} = 0.38$, $1.4$, $3$ TeV, shown in 
Figure~\ref{HProd_CLIC}. We show both the case of unpolarized electron and positron beams (solid lines) and the possibility of using 
beam polarization, which can constitute a strong advantage in searching 
for new physics~\cite{MoortgatPick:2005cw}, assuming for definiteness 
an electron-positron beam polarization$P_{e^-},\,P_{e^+} = -80\%, \, + 30\%$ 
(dashed lines)\footnote{Here, $-100\%$ corresponds to a fully left-handed polarized beam and 
$+100\%$ to a fully right-handed polarized beam.}  in the ballpark of the expected
CLIC operation setup.

As highlighted in Figure~\ref{HProd_CLIC}, the dominant Higgs production mechanism for both $\sqrt{s} = 1.4$ and $3$ TeV is the 
vector boson fusion (VBF) process $e^+ e^- \to H  \nu \nu$.
We also emphasize that the setup $\sqrt{s} = 380$ GeV does not allow to probe high values of $m_H$, and moreover it does not 
yield as many kinematical handles to disentangle the heavy scalar 
signal from SM backgrounds. In the rest of the paper we then focus on $e^+ e^- \to H \nu \nu$ as Higgs production mechanism in CLIC, considering  
$\sqrt{s} = 1.4$ and $3$ TeV as c.o.m.~energies. The
respective projected integrated luminosities we consider are $\mathcal{L} = $ 1500 fb$^{-1}$ and 2000 fb$^{-1}$~\cite{CLIC:2016zwp}.  
In all our subsequent analyses, we simulate CLIC production of the new scalar $H$ via $e^+ e^- \to H \nu \nu$ using
{\sc Madgraph$\_$aMC@NLO}~\cite{Alwall:2014hca} with a subsequent decay into the relevant final state, 
and assuming electron and positron polarized beams with $P_{e^-},\,P_{e^+} = -80\%, \, + 30\%$ in all our analyses.
We then shower/hadronise our events with {\sc Pythia 8.2}~\cite{Sjostrand:2014zea} and use {\sc Delphes}~\cite{deFavereau:2013fsa} 
for a simulation of the detector performance with the Delphes Tune for CLIC studies~\cite{UlrikeGitHub,AlipourTehrani:2254048} (see also~\cite{Potter:2016pgp}).

\section{Searching for heavy scalars in $VV$ final states with $\sqrt{s} = 3$ TeV}
\label{subsec:VV}
   
We examine here the CLIC potential to search for new scalars via decays into EW gauge bosons $H \to V V$ ($V = W^{\pm}, \,Z$). 
We focus on leptonic final states $H \to 4\ell$ in Section~\ref{Sec4l} and $H \to 2\ell\, 2\nu$ in Section~\ref{Sec2l2nu}, and leave
hadronic final states (requiring a more involved analysis, but being very promising due to the 
large branching fraction and the clean environment of CLIC)
for a future analysis. We restrict our analysis to a CLIC c.o.m.~energy $\sqrt{s} = 3$ TeV for our $V V$ studies, 
as our results will show that the projected sensitivity for $\sqrt{s} = 1.4$ TeV would not be competitive 
with that of HL-LHC. In addition, for the $H \to 2\ell\, 2\nu$ final state analysis of Section~\ref{Sec2l2nu}, 
we focus on the $H \to W^+W^- \to 2\ell\, 2\nu$ signal decay channel:
we have found that the projected sensitivity of this channel is significantly larger than the one that can be achieved 
for the $H \to Z Z \to 2\ell\, 2\nu$ signal channel, and thus disregard the latter in the following.

\begin{table}[h]

\begin{center}

\begin{tabular}{l| c | c | c | c | }

\textbf{$\sqrt{s} = 3$ TeV} & $\sigma^{300}_S$ & $\sigma^{600}_S$  & $ \sigma^{900}_S$ & $\sigma^{ZZ\nu\nu}_B$\\
\hline 
&  &  &  & \\ [-2ex]
Event selection& 0.711 & 0.388 & 0.107 & 0.303 \\ [0.5ex]
{\underline {\bf $H\to 4\ell$ selection}}& &  &  &   \\ [0.5ex]
$\chi(m_{\ell_a\ell_b},m_{\ell_c\ell_d}) < 1$& 0.631  & 0.351  & 0.096  & 0.232 \\ [0.5ex]
SR$_{300}$& 0.621  &   &  & 0.017 \\ [0.5ex]
SR$_{600}$&   & 0.319  &  & 0.0053 \\ [0.5ex]
SR$_{900}$&   &   & 0.075 & 0.0016 \\ [0.5ex]
\hline
\end{tabular}
\caption{\small 3 TeV CLIC cross section (in fb) for signal (for $m_H = 300$, $600$, $900$ GeV respectively) and the dominant SM
background $\sigma^{ZZ\nu\nu}_B$ at different stages in the event selection and in the signal region (SR)
for $m_H = 300$, $600$, $900$ GeV respectively (see text for details).
}
\label{Table_3000_VV_4l}
\end{center}
\end{table}

\vspace{-4mm}

\subsection{$H \to 4 \ell$}    
\label{Sec4l}

\begin{figure}[h]

\begin{center}
\includegraphics[width=0.70\textwidth]{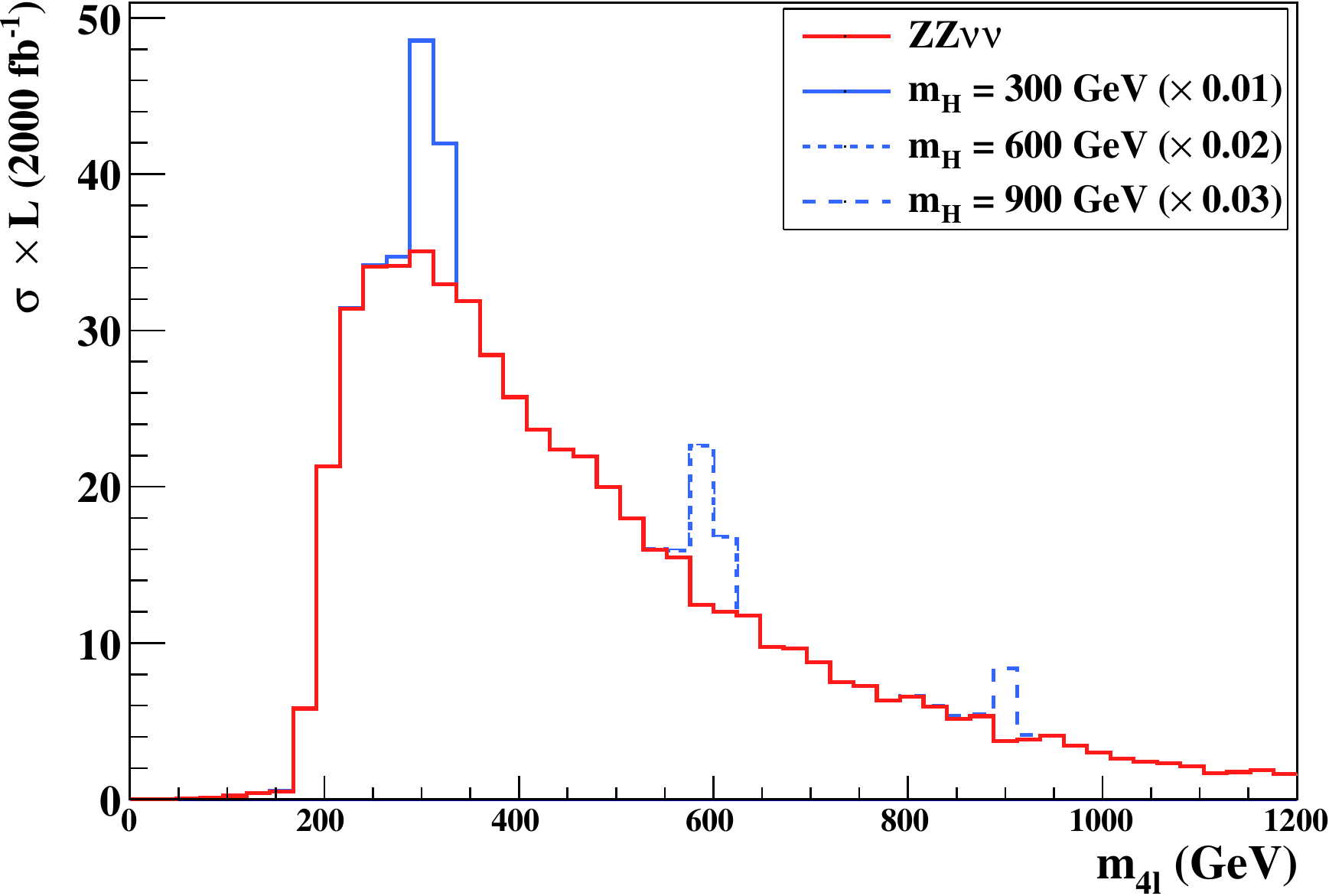}

\caption{\small $m_{4\ell}$ distribution  (with the vertical axis corresponding to the number of events for an integrated luminosity 
$\mathcal{L} = 2000$ fb$^{-1}$) for the signal $e^+ e^- \to H \nu\nu$ ($H \to ZZ \to 4 \ell$) with $m_H = 300$ GeV (solid blue), $600$ GeV (dotted blue), 
$900$ GeV (dashed blue) and the dominant SM background $e^+ e^- \to Z Z \nu\nu\, (ZZ \to 4 \ell)$ (red), 
for $\sqrt{s} = 3$ TeV CLIC.}
\label{CLIC_ZZ4l_30}
\end{center}

\end{figure}

The main SM backgrounds for heavy scalar production (in VBF) and subsequent decay $H \to Z Z \to 4 \ell$ are the 
SM Higgs production $e^+ e^- \to h \nu\nu$ ($h \to 4 \ell$) and the EW processes $e^+ e^- \to Z Z \to 4 \ell$,
$e^+ e^- \to W^+ W^- Z \to 4\ell \,2\nu$, $e^+ e^- \to Z Z \nu \nu \,(Z Z \to 4\ell)$. 
As initial event selection, we require four reconstructed leptons within the detector acceptance 
region ($\left|\eta_{\ell} \right| \leq 2.54$ for electrons and muons), 
yielding two same-flavour lepton pairs. In case of multiple possible pairings among the four leptons $\ell_{a,b,c,d}$ 
we choose the pairing minimising the function $\chi(m_{\ell_a\ell_b}, m_{\ell_c\ell_d})$
\begin{equation}
\chi = \sqrt{\frac{\left(m_{\ell_a\ell_b} - m_Z \right)^2}{\Delta m_Z^2} + \frac{\left(m_{\ell_c\ell_d} - m_Z \right)^2}{\Delta m_Z^2}} 
\end{equation}
with $m_Z = 91$ GeV and the choice $\Delta m_Z = 12 $ GeV. We then select events for which $\chi < 1$, and define the signal region (SR) as 
the invariant mass window
$m_{4\ell} \in [m_H - 15\,\mathrm{GeV},\, m_H + 12\,\mathrm{GeV}]$. We note that apart from the 
process $e^+ e^- \to Z Z \nu \nu \,(Z Z \to 4\ell)$, the contribution of the SM backgrounds to the signal region is 
negligible\footnote{The SM Higgs and $e^+ e^- \to W^+ W^- Z$ backgrounds are strongly suppressed by the condition $\chi < 1$, 
while the $e^+ e^- \to Z Z$ background is severely reduced by
reconstructing the invariant mass $m_{4\ell}$ at values significantly away from $\sqrt{s} = 3$ TeV.} (less than one event expected for an integrated 
luminosity $\mathcal{L} = 2000\,\, \mathrm{fb}$).  
The cross section of the SM $e^+ e^- \to Z Z \nu \nu \,(Z Z \to 4\ell)$ background and three 
benchmark signal scenarios ($m_H = 300$ GeV, $600$ GeV, $900$ GeV) at various stages in the selection process is shown in Table~\ref{Table_3000_VV_4l}.
We also show the $m_{4\ell}$ invariant mass distribution after event selection 
for the $Z Z \nu\nu$ SM background and the three benchmark signal scenarios in 
Figure~\ref{CLIC_ZZ4l_30}.

\begin{figure}[h]

\begin{center}
\includegraphics[width=0.77\textwidth]{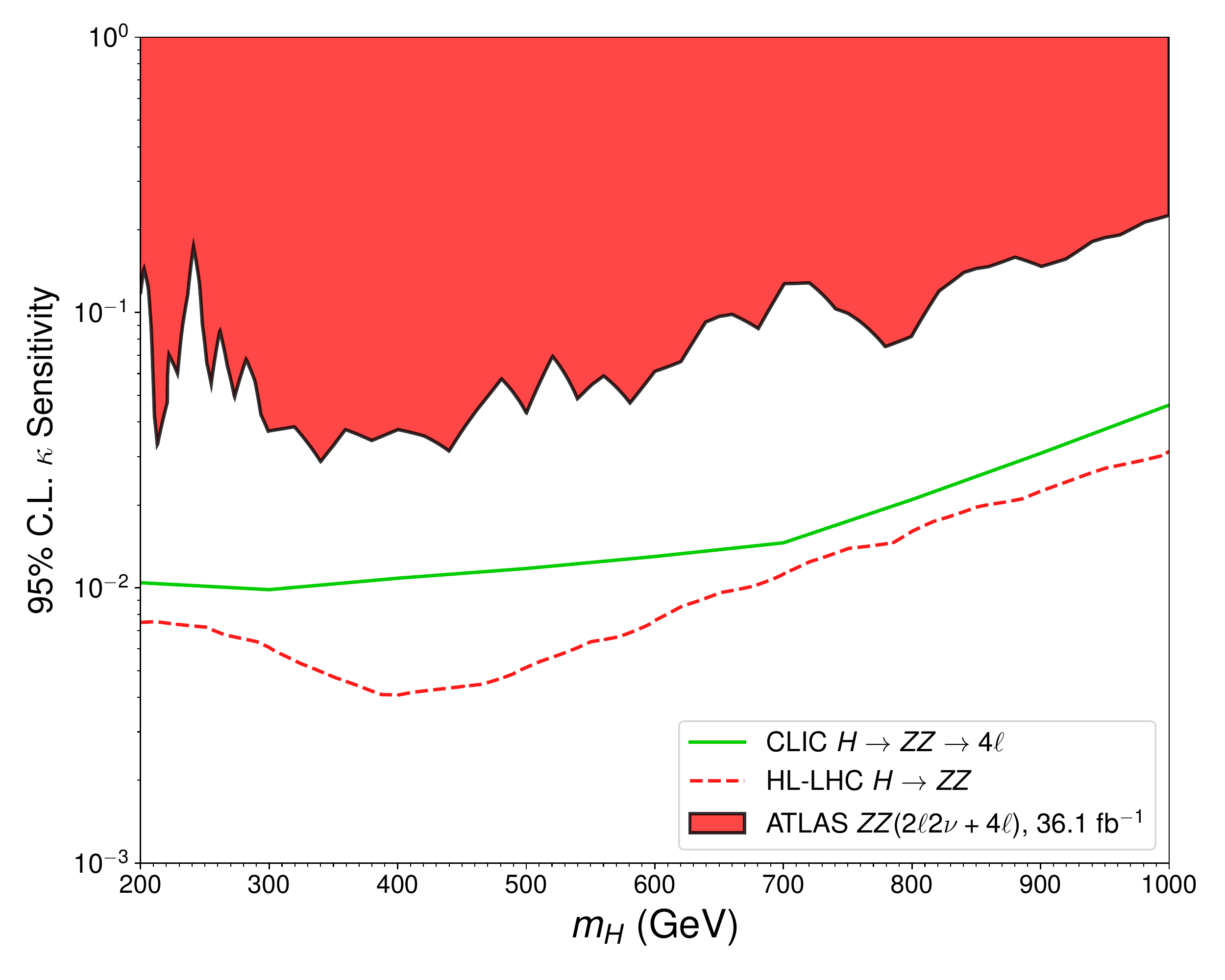}

\caption{\small 95\% C.L. sensitivity to $\kappa = \sigma_S/\sigma^{\mathrm{SM}}_S$ as a function of $m_H$ for $e^+ e^- \to H \nu\nu$ ($H \to ZZ \to 4 \ell$)
at 3 TeV CLIC with $\mathcal{L} = 2000$ fb$^{-1}$ (solid green line). Shown for comparison are the 95\% C.L. excluded region from present ATLAS 
$H \to ZZ$ searches~\cite{Aaboud:2017rel} (red region) and the projected HL-LHC ($13$ TeV, $\mathcal{L} = 3$ ab$^{-1}$) expected 95\% C.L. 
exclusion sensitivity (dashed red line).}
\label{Sensitivity_ZZ_4l_3000}
\end{center}

\end{figure}

From the above analysis, we obtain the projected 95\% C.L. sensitivity reach of $\sqrt{s} = 3$ TeV 
CLIC with 
$\mathcal{L} = 2000\,\, \mathrm{fb}$, in the mass range 
$m_H \in [200\,\mathrm{GeV}, \,1 \,\mathrm{TeV}]$. We perform a likelihood analysis 
based on the number of signal ($s$) and background ($b$) events in the signal region, 
the (Poisson) likelihood function given by
\begin{eqnarray}
L(\kappa) = e^{-(\kappa\, s +\, b)}\, \frac{(\kappa\, s + b)^{n}}{n !}   
\label{likelihood_NS}
\end{eqnarray}
with the number of observed events ($n$) assumed to match the background prediction ($n = b$). 
The signal strength $\kappa \equiv \sigma_S/\sigma^{\mathrm{SM}}_S$ is defined as 
the ratio of the signal cross section to its value assuming SM values (for a given $m_H$) for both 
the production cross section of $H$ and its branching fraction $H \to Z Z$. 
We use the test statistic $Q_{\kappa}$
\begin{eqnarray}
Q_{\kappa} \equiv -2\, \mathrm{Log} \left[\frac{L(\kappa)}{L(0)} \right]\,,
\label{likelihood_1}
\end{eqnarray}
to obtain the 95\% C.L. exclusion sensitivity, given by $Q_{\kappa} = 3.84$.
This is shown in Figure~\ref{Sensitivity_ZZ_4l_3000} (solid green line). For comparison, we show 
the present ($\sqrt{s} = 13$ TeV LHC with $\mathcal{L} = 36.1$ fb$^{-1}$) 
limits on $\kappa$ from ATLAS $H \to ZZ$ searches~\cite{Aaboud:2017rel}, with the SM
(gluon fusion) production cross section for $H$ obtained from~\cite{Heinemeyer:2013tqa}. 
We also show the HL-LHC ($\sqrt{s} = 13$ TeV with $\mathcal{L} = 3$ ab$^{-1}$) projected 95\% C.L. sensitivity 
from a naive $\sqrt{\mathcal{L}}$ scaling w.r.t.~to the present expected exclusion sensitivity from~\cite{Aaboud:2017rel}.
As is apparent from Figure~\ref{Sensitivity_ZZ_4l_3000}, the sensitivity that can be achieved by CLIC in heavy scalar searches 
$H \to Z Z \to 4\ell$ is at best comparable to that of HL-LHC. However, we emphasize that while heavy scalar searches via 
leptonic final states are bound to yield the best sensitivity at the LHC, for CLIC it is expected that hadronic final 
states could surpass the sensitivity of leptonic ones, and a future study in this direction is well worth pursuing.


\vspace{2mm}

As a final remark on the analysis, we stress that for $m_H \gtrsim 1$ TeV the mean separation between the two leptons 
coming from each $Z$ decay $\Delta R \sim 2 m_Z / |\vec{P}_Z| \sim 
4 m_Z / m_H < 0.4$ and our analysis 
(which imposes a lepton isolation $\Delta R^{\mathrm{min}} = 0.5$ from the {\sc Delphes} lepton reconstruction criteria) 
becomes highly inefficient. Gaining sensitivity to higher masses requires decreasing the 
required $\Delta R^{\mathrm{min}}$ lepton isolation (as e.g.~exemplified in~\cite{Aaboud:2017rel}).
Still, it will be shown in Section~\ref{sec:xSM}
that the relevant mass range to consider for the EW phase transition in the scenarios we will analyse is 
$m_H \lesssim 1$ TeV, and the lepton isolation criteria in our analysis are thus well-justified.

\vspace{-4mm}

\subsection{$H \to  2\ell\, 2\nu$}
\label{Sec2l2nu}

The relevant SM backgrounds for $H$ production through $e^+ e^- \to H \nu\nu$ and subsequent decay $H \to W^+ W^- \to 2\ell\, 2\nu$ are:

\vspace{2mm}

\noindent {\it (i)} SM Higgs production through VBF: $e^+ e^- \to h \,\nu \nu$ ($h \to 2\ell\, 2\nu$)

\vspace{1mm}

\noindent {\it (ii)} EW processes yielding a $2\ell \,2\nu$ final state. These include $e^+ e^- \to W^+ W^- (\to 2\ell \,2\nu)$, $e^+ e^- \to Z Z \,(\to 2\ell \,2\nu)$, 
$e^+ e^- \to e^{\pm } \nu W^{\mp} \,(W^{\mp} \to \ell^{\mp} \nu)$, $e^+ e^- \to Z \nu \nu \,(Z \to 2\ell)$, $e^+ e^- \to Z e^+ e^- (Z \to 2\nu)$
(in the last three processes, the states accompanying the produced $W^{\pm}$ or $Z$ boson do not themselves come from a $W^{\pm}$ or $Z$ boson).

\vspace{1mm}

\noindent {\it (iii)} The dominant EW processes yielding a $2\ell \,4\nu$ final state:  
$e^+ e^- \to W^+ W^- \nu \nu \,(W^+ W^-\to 2\ell \,2\nu)$ and $e^+ e^- \to Z Z \nu \nu \,(Z Z \to 2\ell \,2\nu)$ 
(including the case where the initial neutrinos come from an on-shell $Z$ boson).

\vspace{1mm}

\noindent {\it (iv)} We also include the process $e^+ e^- \to \gamma \,2\ell$ (including the case where the two leptons come from an on-shell $Z$ boson). 
%
 
\vspace{2mm} 

For event selection we require two reconstructed leptons $\ell = e, \mu$ in the final state 
with $\left|\eta_{\ell}\right| \leq 2.44$. 
In addition, we require $m_{\ell\ell} \geq 100$ GeV to suppress backgrounds where the two leptons 
are coming from an on-shell $Z$ boson, as well as the SM Higgs background.
In order to subsequently suppress the SM backgrounds, we require $\left|\eta_{\ell}\right| \leq 1.5$ 
(the signal events feature rather central leptons, as opposed to several SM backgrounds) and $1 \leq \Delta R_{\ell\ell} \leq 3.5$.
Finally, we also require $P_{\ell\ell} \leq  500$ GeV. 

\begin{figure}[h]

\begin{center}
\includegraphics[width=0.49\textwidth]{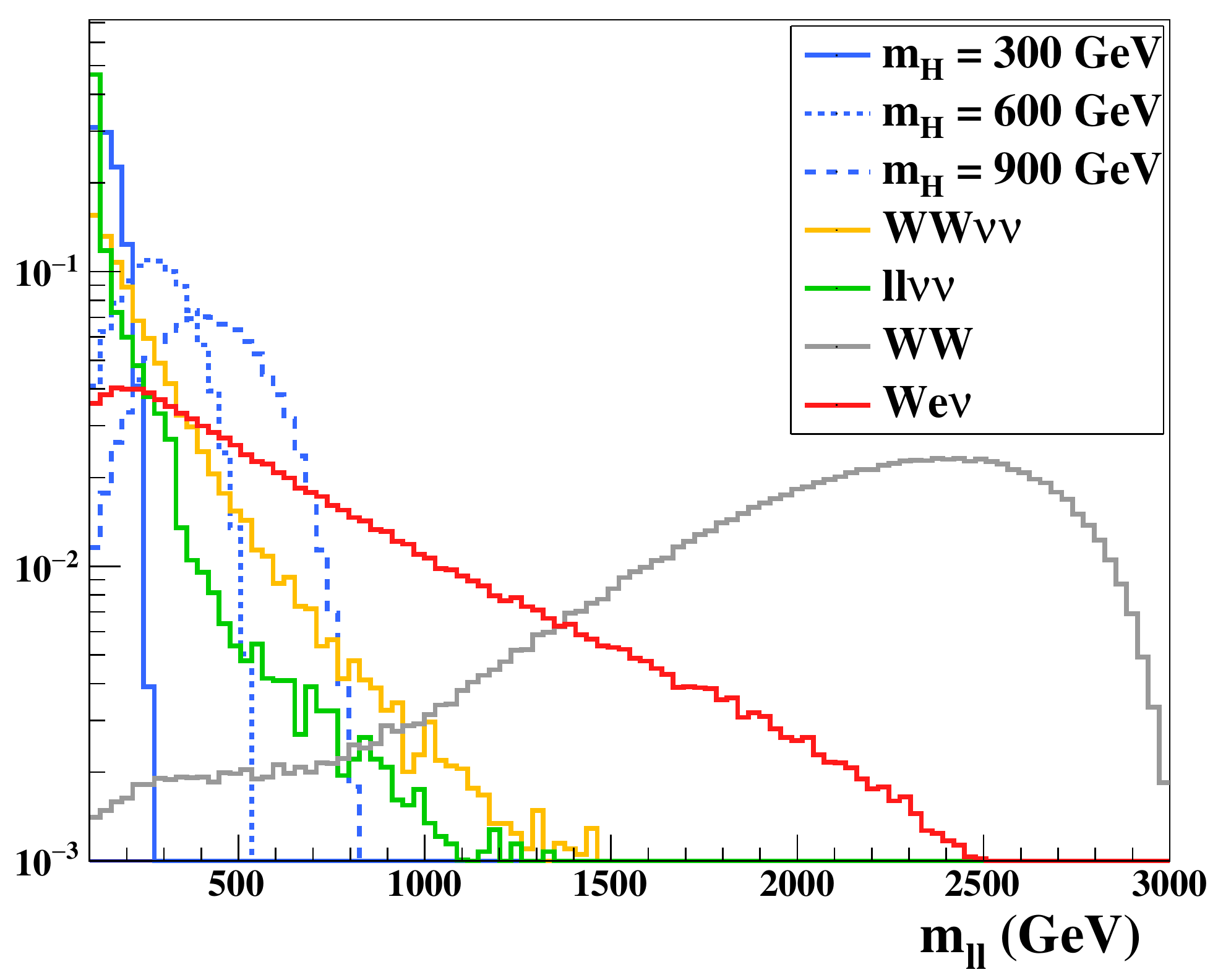}
\includegraphics[width=0.49\textwidth]{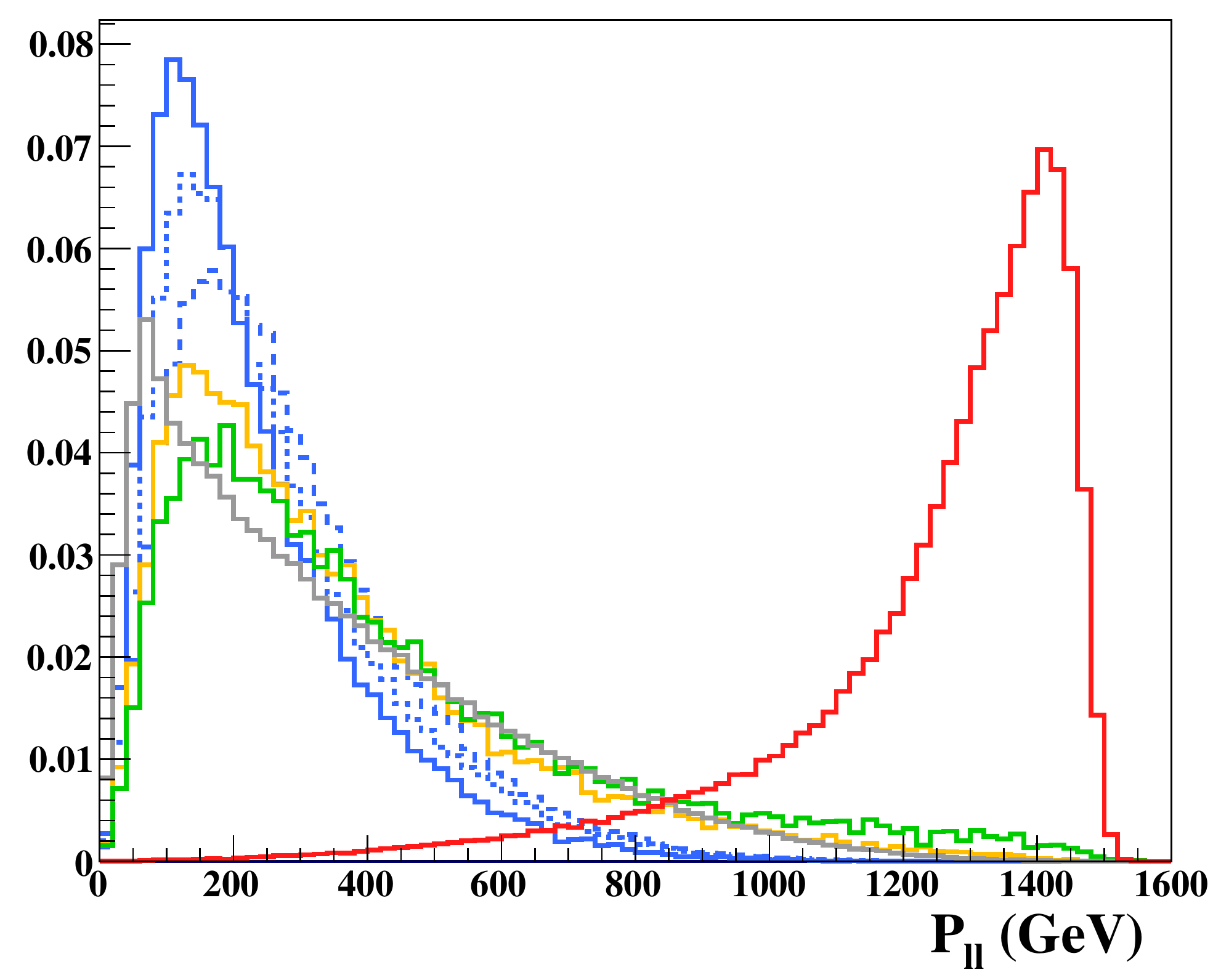}

\caption{\small Normalized kinematic distributions $m_{\ell\ell}$ (left) and $P_{\ell\ell}$ (right)
for the signal $e^+ e^- \to H \nu\nu$ ($H \to WW \to 2\ell2\nu$) with $m_H = 300$ GeV (solid blue), $600$ GeV (dotted blue), 
$900$ GeV (dashed blue) and SM backgrounds $e^+ e^- \to WW$ (grey), $e^+ e^- \to e^{\pm }\nu W^{\mp}$ (red), 
$e^+ e^- \to WW\nu\nu$ (yellow) and $e^+ e^- \to \ell\ell \nu\nu$ (green), for $\sqrt{s} = 3$ TeV CLIC.}
\label{CLIC_WW_30_1}
\end{center}

\end{figure}

\begin{figure}[t]

\begin{center}

\includegraphics[width=0.49\textwidth]{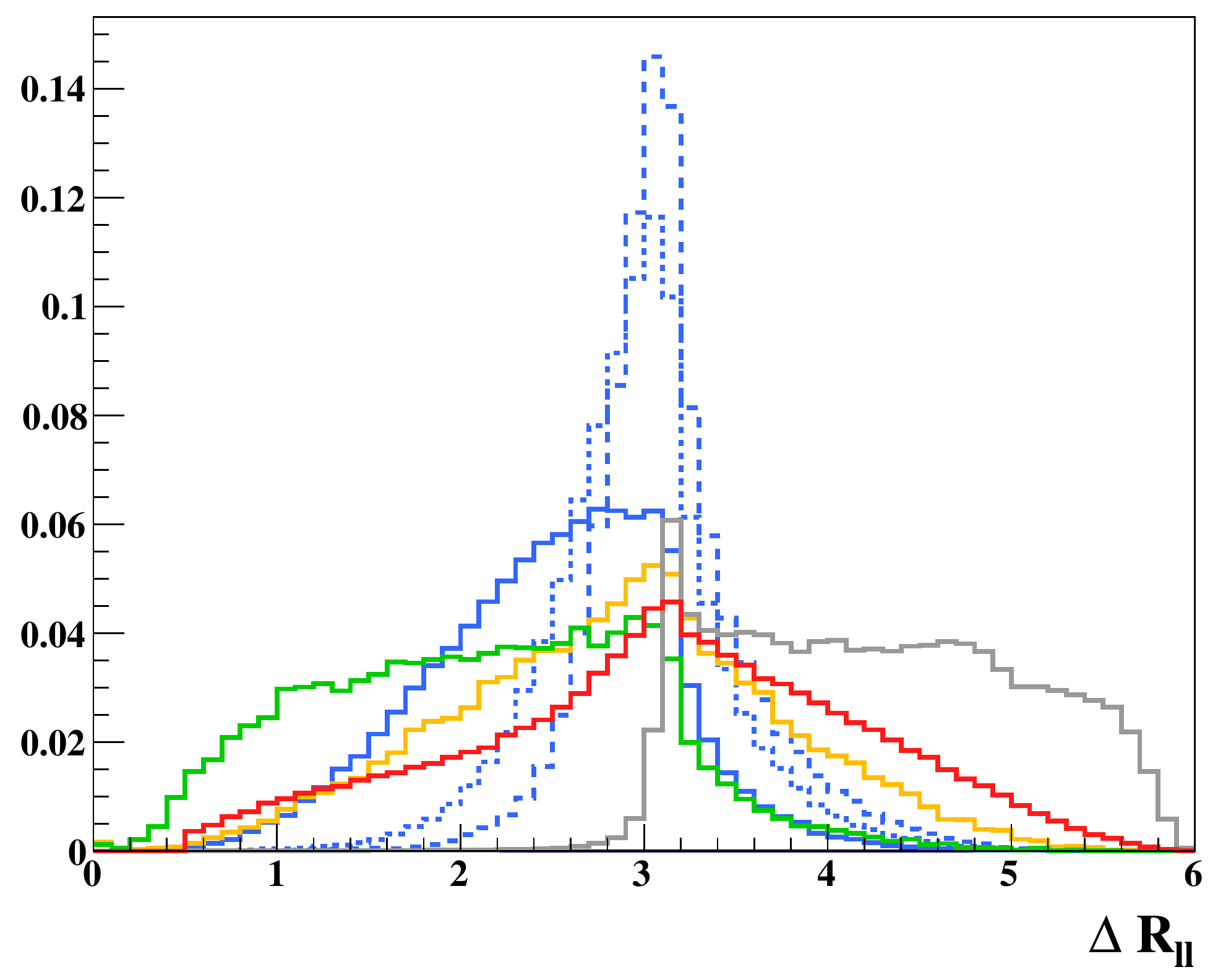}
\includegraphics[width=0.49\textwidth]{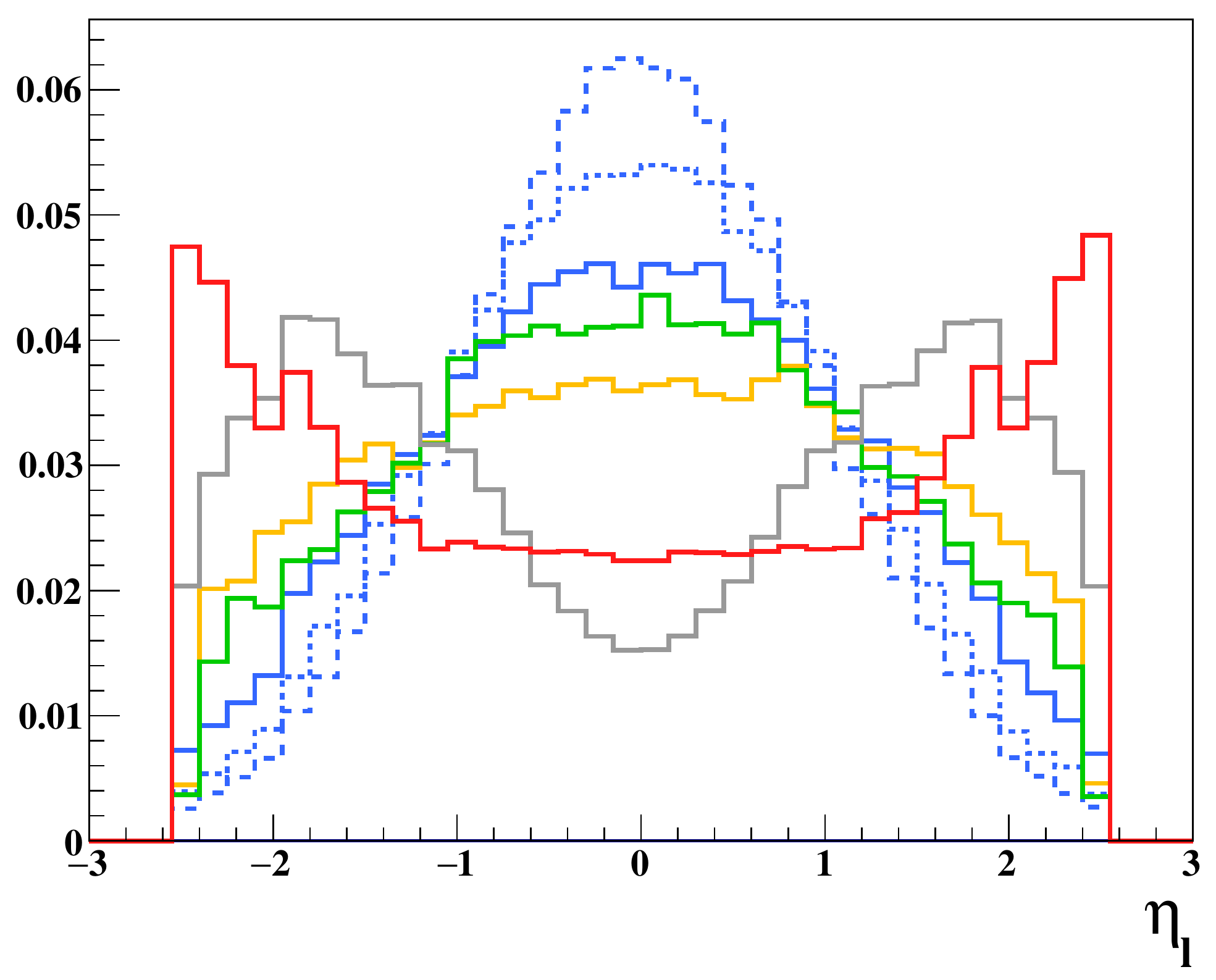}

\caption{\small Normalized kinematic distribution $\Delta R_{\ell\ell}$ (left) and $\eta_\ell$ (right)
for the signal $e^+ e^- \to H \nu\nu$ ($H \to WW \to 2\ell2\nu$) with $m_H = 300$ GeV (solid blue), $600$ GeV (dotted blue), 
$900$ GeV (dashed blue) and SM backgrounds $e^+ e^- \to WW$ (grey), $e^+ e^- \to e^{\pm }\nu W^{\mp}$ (red), 
$e^+ e^- \to WW\nu\nu$ (yellow) and $e^+ e^- \to \ell\ell \nu\nu$ (green), for $\sqrt{s} = 3$ TeV CLIC.}
\label{CLIC_WW_30_2}
\end{center}

\end{figure}

After the above selection cuts, the background from the SM Higgs becomes completely negligible. In addition, the $m_{\ell\ell}$ spectrum for the 
backgrounds $e^+ e^- \to \gamma \,2\ell$ and $e^+ e^- \to Z e^+ e^- (Z \to 2\nu)$ after the selection cuts features $m_{\ell\ell} \gtrsim 2$ TeV, which leads to 
a negligible overlap with the signal region domain (discussed below). 
In the following, we then consider as dominant backgrounds the processes 
$e^+ e^- \to W^+ W^- (\to 2\ell \,2\nu)$, $e^+ e^- \to e^{\pm } \nu W^{\mp} \,(W^{\mp} \to \ell^{\mp} \nu)$, 
$e^+ e^- \to \ell\ell \,\nu\nu$ (with the final states not coming from $W$ boson(s)) and 
$e^+ e^- \to W^+ W^- \nu \nu \,(W^+ W^-\to 2\ell \,2\nu)$.
The (normalized) 
$m_{\ell\ell}$, $P_{\ell\ell}$, $\eta_{\ell}$ and $\Delta R_{\ell\ell}$ kinematic distributions after event selection and imposing 
$m_{\ell\ell} \geq 100$ GeV are shown 
in Figures~\ref{CLIC_WW_30_1}-\ref{CLIC_WW_30_2}. 

\noindent We define the signal region SR as:
\begin{equation}
\mathrm{max}(100\,\,\mathrm{GeV},\,C - \Delta) \leq m_{\ell \ell} \leq C + \Delta \quad, \quad 
\left\lbrace  \begin{array}{l}
C(m_H) = 0.457 \times m_H - 15 \,\mathrm{GeV}\\
\Delta (m_H) = 0.264 \times m_H - 6.5 \,\mathrm{GeV}
\end{array}  \right.
\label{SR_WW_mll}
\end{equation}
which we obtain from an approximate fit to the $m_H$-dependence 
of the $m_{\ell\ell}$ distribution's peak (median) and width ($1.5 \times$variance) 
for our signal samples after the event and cut-flow selection discussed above.
The cross sections for the relevant backgrounds and signal benchmarks with $m_H = 300$ GeV, $600$ GeV, $900$ GeV
after event selection, the subsequent cut-flow and the final signal region selection are given in Table~\ref{Table1_3000}. 

\begin{table}[h]

\begin{center}

\begin{tabular}{l| c | c | c | c | c | c | c | c | c |}

\textbf{$\sqrt{s} = 3$ TeV} & $\sigma^{300}_S$ & $\sigma^{600}_S$  & $ \sigma^{900}_S$ & $\sigma^{WW}_B$ & $\sigma^{We\nu}_B$  
& $\sigma^{\ell\ell\nu\nu}_B$  & $\sigma^{WW\nu\nu}_B$\\
\hline 
&  &  &  & &  & & \\ [-2ex]
Event selection& 18.9 & 9.3 & 6.0  & 11.3 & 261 & 199 & 10.6 \\ [0.5ex]
{\underline {\bf $H\to WW$ selection}}& &  &  &  &  &  &  \\ [0.5ex]
$m_{\ell\ell} \geq 100$ GeV& 13.1 & 9.0  & 5.95  & 11.2 & 248 & 15.2 & 4.63  \\ [0.5ex]
$\left|\eta_{\ell}\right| \leq 1.5$, $1 \leq \Delta R_{\ell\ell} \leq 3.5$& 7.92 & 6.26 & 4.45 & 2.56 & 31.3  & 7.35 &  2.93 \\ [0.5ex]
$P_{\ell\ell} \leq  500$ GeV& 7.88   & 5.98  & 4.04 & 1.90 & 0.51  & 6.56 & 2.39  \\ [0.5ex]
SR$_{300}$& 6.90  &   &  & 0.043  &  0.138 & 4.79 & 1.32  \\ [0.5ex]
SR$_{600}$&   &  5.41 &  & 0.154 & 0.226  & 4.65 & 2.03  \\ [0.5ex]
SR$_{900}$&   &   & 3.57 & 0.229 & 0.152  & 2.19 & 1.28  \\ [0.5ex]
\hline
\end{tabular}
\caption{\small 3 TeV CLIC cross section (in fb) for signal (for $m_H = 300$, $600$, $900$ GeV respectively) and SM
backgrounds $\sigma^{WW}_B$, $\sigma^{We\nu}_B$, $\sigma^{\ell\ell\nu\nu}_B$, $\sigma^{WW\nu\nu}_B$ 
at different stages in the event, cut-flow and $H\to WW \to 2\ell\,2\nu$ signal region (SR) selection
(see text for details).
}
\label{Table1_3000}
\end{center}
\end{table}

\begin{figure}[h]

\begin{center}
\includegraphics[width=0.75\textwidth]{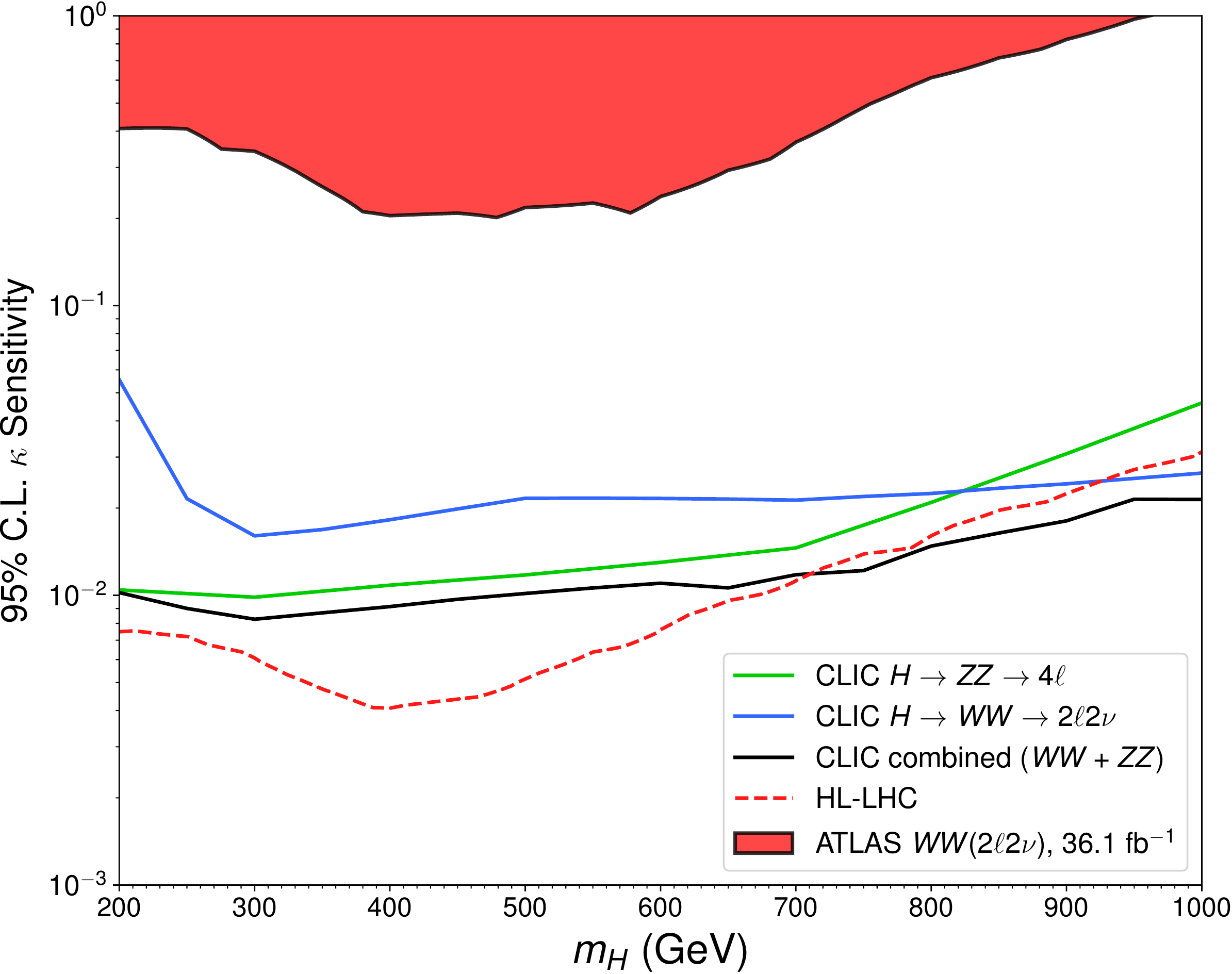}

\caption{\small 95\% C.L. sensitivity to $\kappa = \sigma_S/\sigma^{\mathrm{SM}}_S$ as a function of $m_H$ for $e^+ e^- \to H \nu\nu$ ($H \to WW \to 2\ell2\nu$)
at 3 TeV CLIC with $\mathcal{L} = 2000$ fb$^{-1}$ (solid blue line), together with the sensitivity for $H \to ZZ \to 4\ell$ 
from Figure~\ref{Sensitivity_ZZ_4l_3000} (solid green line) and the combined sensitivity (solid black line).
Shown for comparison are the 95\% C.L. excluded region from present ATLAS 
$H \to WW$ searches~\cite{Aaboud:2017gsl} (red region) and the projected HL-LHC ($13$ TeV, $\mathcal{L} = 3$ ab$^{-1}$) 95\% C.L. 
sensitivity reach (dashed red line), dominated by $H \to ZZ$ searches (see section~\ref{Sec4l}).}
\label{Sensitivity_WW_4l_3000}
\end{center}

\end{figure}

Assuming $\mathcal{L} = 2000$ fb$^{-1}$, we show the projected 95\% C.L. sensitivity reach of the 
$e^+ e^- \to H \nu\nu$ ($H \to WW \to 2\ell2\nu$) search at $\sqrt{s} = 3$ TeV CLIC in Figure~\ref{Sensitivity_WW_4l_3000}, following the 
likelihood analysis already employed in section~\ref{Sec4l}. 
We note the partial loss of sensitivity for $m_H < 300$ GeV, as the $m_{\ell\ell}$ distribution
for the signal mainly lies under the $Z$-peak of the $\ell\ell \nu\nu$ SM background, as can be inferred from Figure~\ref{CLIC_WW_30_1}.
Figure~\ref{Sensitivity_WW_4l_3000} also shows the CLIC sensitivity reach in $\kappa = \sigma_S/\sigma^{\mathrm{SM}}_S$ 
from the combination of the $H \to WW \to 2\ell2\nu$ and $H \to ZZ \to 4\ell$ (see section~\ref{Sec4l}) signal channels.
For the sake of comparison, we show as well the present LHC limits for $H \to WW \to 2\ell2\nu$ searches 
from ATLAS~\cite{Aaboud:2017gsl} ($\sqrt{s} = 13$ TeV LHC with $\mathcal{L} = 36.1$ fb$^{-1}$), together with the 
projected 95\% C.L. sensitivity reach in $\kappa$ of ($\sqrt{s} = 13$ TeV) HL-LHC, which is essentially dominated by the 
$H \to ZZ$ searches (and thus corresponds to that shown in Figure~\ref{Sensitivity_ZZ_4l_3000}). Figure~\ref{Sensitivity_WW_4l_3000} highlights that 
$H \to VV$ searches at CLIC in the leptonic channels reach a comparable sensitivity to that of HL-LHC. 

\vspace{-6mm}  

\section{Searching for heavy scalars in $hh$ final states}
\label{subsec:hh}

We now turn to explore the CLIC sensitivity to new scalars through resonant di-Higgs signatures $H \to hh$.
We focus on the $hh \to b\bar{b}b\bar{b}$ final state, which
has the largest branching fraction while it does not suffer from the 
very large QCD background one has to face in the LHC environment \cite{Dolan:2012rv, deLima:2014dta}.
We will show in the following that resonant di-Higgs searches at CLIC constitute a very sensitive 
probe of the existence of new scalars. In Section~\ref{Sechh3TeV} we analyse the 
$\sqrt{s} = 3$ TeV CLIC prospects, and discuss those for $\sqrt{s} = 1.4$ TeV in Section~\ref{Sechh14TeV}.

\subsection{$\sqrt{s} = 3$ TeV}  
\label{Sechh3TeV} 
 
The dominant backgrounds to the $e^+ e^- \to H \nu\nu$ ($H \to hh \to 4 b$) process at CLIC are from EW  
(including the SM non-resonant di-Higgs production contribution, 
on which we will comment in Section~\ref{sec:xSM}) and QCD processes yielding a 
$4b + 2\nu$ final state. We reconstruct jets (within {\sc Delphes}) with 
{\sc Fastjet}~\cite{Cacciari:2011ma}, using the Valencia clustering algorithm~\cite{Boronat:2014hva} (particularly well-suited for jet reconstruction in 
high energy $e^+ e^-$ colliders) in exclusive mode with $R = 0.7$ and $N = 4$ (number of jets). 
We perform our analysis for two different $b$-tagging working points 
within the CLIC Delphes Tune, corresponding respectively to a 70\% and 90\% $b$-tagging 
efficiency\footnote{For the 90\% $b$-tagging working point, the background contribution from events with $c$-jets which are mis-identified as $b$-jets 
ceases to be negligible and should be considered in an exhaustive study. Nevertheless, 
the ratio of $b$-tagging efficiency to $c$-jet mistag rate is in this case $\sim 0.2$ (and backgrounds with 
mis-identified $c$-jets need to contain at least two of those), such that events with 
mis-identified jets are still subdominant, and we will not consider them here.}. In each case,  
%
%
%
we select events with 4 $b$-tagged jets, which are subsequently paired into two 125 GeV Higgs candidates by 
minimizing
\begin{equation}
\chi = \sqrt{\frac{\left(m_{b_1 b_2} - \overline{m_h} \right)^2}{\Delta_h^2} + \frac{\left(m_{b_3 b_4} - \overline{m_h} \right)^2}{\Delta_h^2}} 
\end{equation}
where $\overline{m_h} = 102$ GeV and $\Delta_h = 30$ GeV are obtained from an approximate fit to the signal simulation (we note that 
the average Higgs mass $\overline{m_h}$ is somewhat lower than the truth value $m_h = 125$ GeV as a result of the jet reconstruction process).
We then select events with two SM Higgs candidates by requiring 
$\chi < 1$. 

\begin{figure}[h]
\begin{center}
\includegraphics[width=0.95\textwidth]{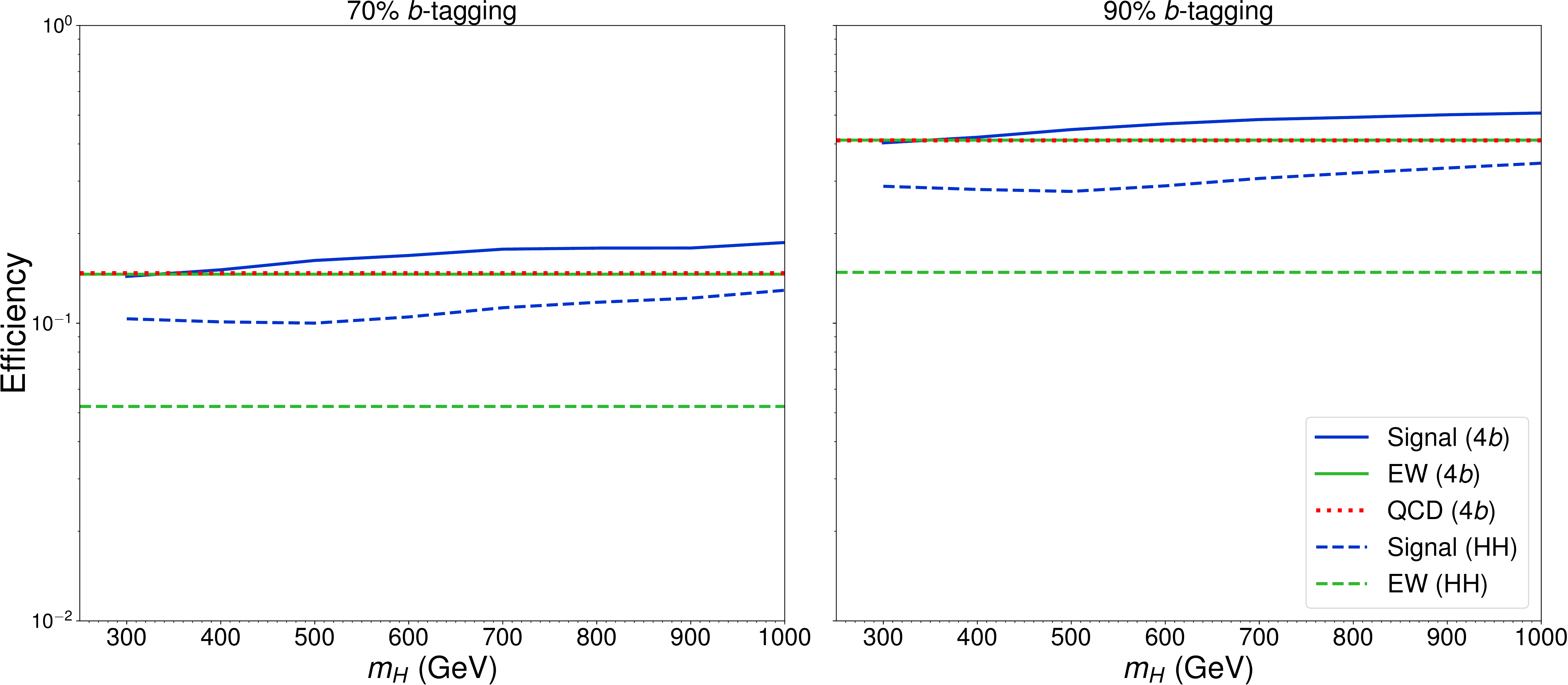}

\caption{\small Signal (blue), EW background (green) and QCD background (red) 
efficiency after $b$-tagging ($4b$, solid/dotted) and SM Higgs candidate selection (HH, dashed) as a function of 
$m_H$ (see text for details).}
\label{Efficiency_hh_3000}
\end{center}
\end{figure}

In Figure~\ref{Efficiency_hh_3000} we show the signal efficiency after $b$-tagging and SM Higgs candidate selection (HH) as a function of 
$m_H$, together with the corresponding background (both EW and QCD) efficiencies (independent of $m_H$).  
After the SM Higgs candidate selection, the efficiency for the QCD background drops dramatically ($\sim 7\times 10^{-5} $ for a 70\% $b$-tagging efficiency 
and $\sim 2\times 10^{-3}$ for a 90\% $b$-tagging efficiency), such that the only relevant SM background is from the EW processes discussed above. 

\begin{table}[h]
\begin{center}
\begin{tabular}{l| c | c | c | c | c |}
\textbf{$\sqrt{s} = 3$ TeV} & $\sigma^{300}_S$ & $\sigma^{600}_S$  & $ \sigma^{900}_S$ & $\sigma^{EW}_B$ & $\sigma^{QCD}_B$\\
\hline 
&  &  &  & &\\ [-2ex]
Event selection   ($70$\% $b$-tagging) & 12.85 & 8.52  & 5.19 & 0.407 & 0.048 \\ [0.5ex]
{\underline {\bf $H\to hh$ selection}}& &  &  &   &\\ [0.5ex]
$\chi(m_{b_1 b_2},m_{b_3 b_4}) < 1$& 9.26  & 5.29  & 3.52  & 0.146  & $< 10^{-3} $ \\ [0.5ex]
SR$_{300}$& 8.99  &   &  & 0.0444 & -\\ [0.5ex]
SR$_{600}$&   & 4.80  &  & 0.0236 & -\\ [0.5ex]
SR$_{900}$&   &   & 3.03  & 0.0098& -\\ [0.5ex]
\hline
\end{tabular}
%

\vspace{5mm}

\begin{tabular}{l| c | c | c | c | c |}
\textbf{$\sqrt{s} = 3$ TeV} & $\sigma^{300}_S$ & $\sigma^{600}_S$  & $ \sigma^{900}_S$ & $\sigma^{EW}_B$ & $\sigma^{QCD}_B$\\
\hline 
&  &  &  & &\\ [-2ex]
Event selection   ($90$\% $b$-tagging) & 36.09 & 23.58  & 14.56 & 1.14 & 0.136  \\ [0.5ex]
{\underline {\bf $H\to hh$ selection}}& &  &  &   &\\ [0.5ex]
$\chi(m_{b_1 b_2},m_{b_3 b_4}) < 1$& 25.80  & 14.60  & 9.64  &  0.413 & $< 10^{-3} $ \\ [0.5ex]
SR$_{300}$& 25.01  &   &  & 0.126 & -\\ [0.5ex]
SR$_{600}$&   & 13.32  &  & 0.063 & -\\ [0.5ex]
SR$_{900}$&   &   & 8.25 & 0.028 & -\\ [0.5ex]
\hline
\end{tabular}
\caption{\small UP: \small 3 TeV CLIC cross section (in fb) for signal (for $m_H = 300$, $600$, $900$ GeV respectively) and SM
backgrounds for a $b$-tagging efficiency of $70$\%, at different stages in the event selection and in the signal region (SR) 
for $m_H = 300$, $600$, $900$ GeV respectively (see text for details).
DOWN: Same as above, for a $b$-tagging efficiency of $90$\%.
}
\label{Table_3000_hh_4b}
\end{center}
\end{table}
We define the Signal Region (SR) as 
\begin{equation}
m_{4b} \in [C - \Delta, C + \Delta] \quad, \quad 
\left\lbrace  \begin{array}{l}
C(m_H) = 0.96 \times m_H - 45 \,\mathrm{GeV}\\
\Delta (m_H) = 0.05 \times m_H + 40 \,\mathrm{GeV}
\end{array}  \right.
\label{SR_hh}
\end{equation}
with both $C(m_H)$ and $\Delta (m_H)$ extracted from a fit to the signal simulation. 
The cross section of three 
benchmark signal scenarios ($m_H = 300$ GeV, $600$ GeV, $900$ GeV) and the SM backgrounds 
at various stages in the selection process is shown in Table~\ref{Table_3000_hh_4b}-UP (for a $b$-tagging efficiency of $70$\%)
and~\ref{Table_3000_hh_4b}-DOWN (for a $b$-tagging efficiency of $90$\%).

\vspace{2mm}

From the above analysis, we obtain the projected 95\% C.L. sensitivity reach of $\sqrt{s} = 3$ TeV 
CLIC ($\mathcal{L} = 2000\,\, \mathrm{fb}$) for $H \to hh \to b\bar{b}b\bar{b}$ in the mass range 
$m_H \in [300\,\mathrm{GeV}, \,1 \,\mathrm{TeV}]$ by performing a likelihood analysis, 
with a likelihood function and test statistic given respectively by~\eqref{likelihood_NS} and~\eqref{likelihood_1}.
Here the signal strength $\kappa$ is defined as $\kappa \equiv \sigma_S/\sigma^{\mathrm{SM}}_S \times \mathrm{BR}(H\to hh)$
(with $\sigma_S/\sigma^{\mathrm{SM}}_S$ the ratio of the production cross section of $H$ to its SM value).
The resuls of this section are summarized in Figure~\ref{Sensitivity_H_hh_3000}, and discussed 
in detail in the following section~\ref{Sechh14TeV} together with those obtained for $\sqrt{s} = 1.4$ TeV.

\subsection{$\sqrt{s} = 1.4$ TeV}  
\label{Sechh14TeV} 

We now repeat the above analysis for a CLIC c.o.m. energy $\sqrt{s} = 1.4$ TeV with $\mathcal{L} = 1.5$ ab$^{-1}$. 
The cross sections for the signal (for $m_H = 300$ GeV, $600$ GeV, $900$ GeV) and the SM backgrounds 
are shown in Table~\ref{Table_1400_hh_4b},
with the signal region being defined as in the analysis from section~\ref{Sechh3TeV} and given by eq.~\eqref{SR_hh}.
\begin{table}[h]
\begin{center}
\begin{tabular}{l| c | c | c | c | c |}
\textbf{$\sqrt{s} = 1.4$ TeV} & $\sigma^{300}_S$ & $\sigma^{600}_S$  & $ \sigma^{900}_S$ & $\sigma^{EW}_B$ & $\sigma^{QCD}_B$\\
\hline 
&  &  &  & &\\ [-2ex]
Event selection   ($70$\% $b$-tagging) & 6.18 & 2.17  & 0.456 & 0.140 & 0.039 \\ [0.5ex]
{\underline {\bf $H\to hh$ selection}}& &  &  &   &\\ [0.5ex]
$\chi(m_{b_1 b_2},m_{b_3 b_4}) < 1$ & 4.61  & 1.36 & 0.306  & 0.052  & $< 10^{-3} $ \\ [0.5ex]
SR$_{300}$& 4.50 &   &  & 0.022 & -\\ [0.5ex]
SR$_{600}$&   &  1.24 &  & 0.0068 & -\\ [0.5ex]
SR$_{900}$&   &   & 0.263  & 0.0014 & -\\ [0.5ex]
\hline
\end{tabular}

\vspace{5mm}

\begin{tabular}{l| c | c | c | c | c |}
\textbf{$\sqrt{s} = 1.4$ TeV} & $\sigma^{300}_S$ & $\sigma^{600}_S$  & $ \sigma^{900}_S$ & $\sigma^{EW}_B$ & $\sigma^{QCD}_B$\\
\hline 
&  &  &  & &\\ [-2ex]
Event selection   ($90$\% $b$-tagging) & 17.25  & 5.88  & 1.26  & 0.385  & 0.108  \\ [0.5ex]
{\underline {\bf $H\to hh$ selection}}& &  &  &   &\\ [0.5ex]
$\chi(m_{b_1 b_2},m_{b_3 b_4}) < 1$& 12.85  & 3.64  & 0.843  & 0.143  & $< 10^{-3} $ \\ [0.5ex]
SR$_{300}$& 12.51  &   &  & 0.059 & -\\ [0.5ex]
SR$_{600}$&   & 3.32  &  & 0.018 & -\\ [0.5ex]
SR$_{900}$&   &   & 0.725 & 0.0042 & -\\ [0.5ex]
\hline
\end{tabular}
\caption{\small UP: 1.4 TeV CLIC cross section (in fb) for signal (for $m_H = 300$, $600$, $900$ GeV respectively) and SM
backgrounds for a $b$-tagging efficiency of $70$\%, at different stages in the event selection and in the signal region (SR) 
for $m_H = 300$, $600$, $900$ GeV respectively (see text for details). DOWN: Same as above, for a $b$-tagging efficiency of $90$\%.
}
\label{Table_1400_hh_4b}
\end{center}
\end{table}

\begin{figure}[h]

\begin{center}
\includegraphics[width=0.75\textwidth]{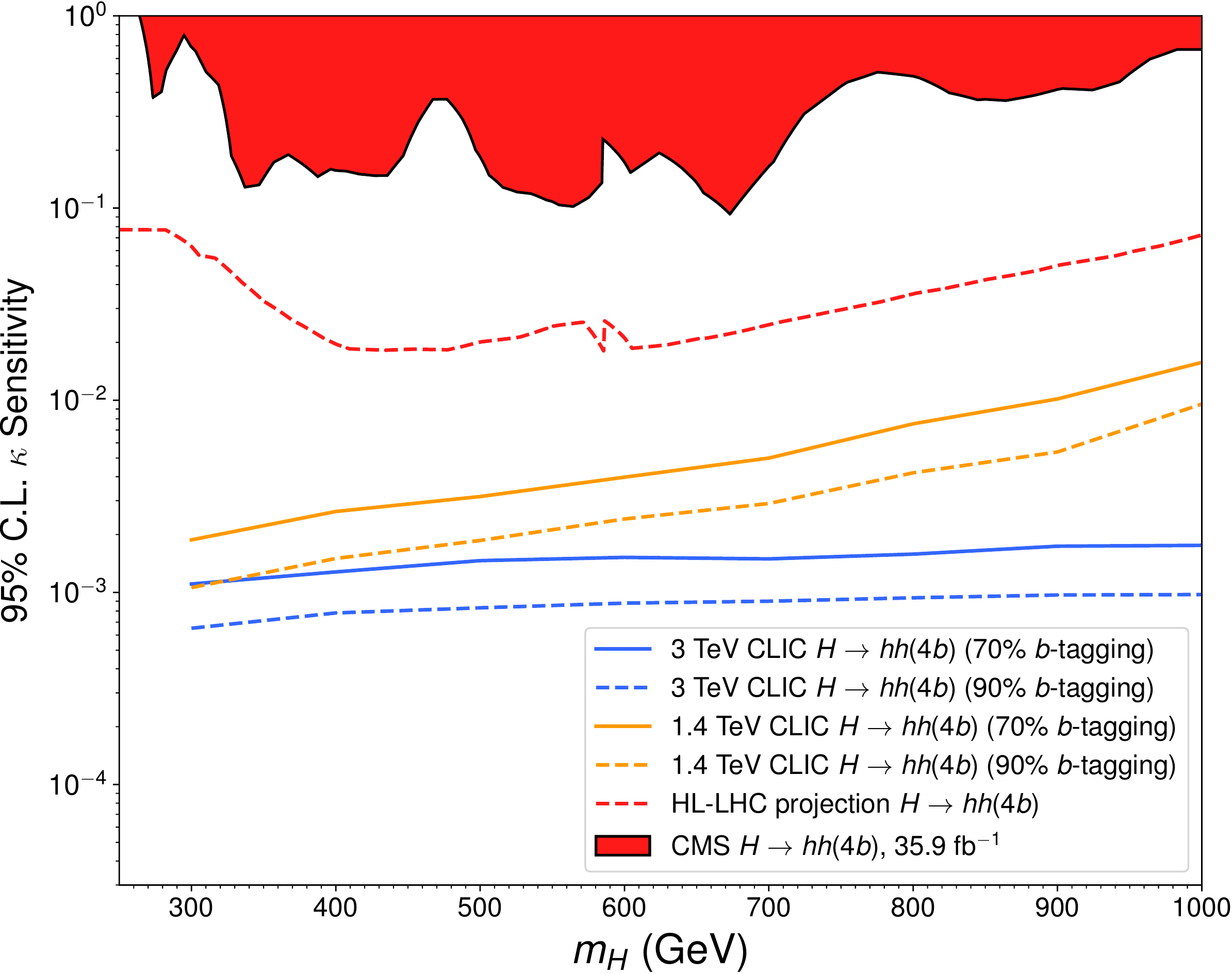}

\caption{\small CLIC 95\% C.L. sensitivity to $\kappa = \sigma_S/\sigma^{\mathrm{SM}}_S \times \mathrm{BR}(H\to hh)$ as a 
function of $m_H$ for $e^+ e^- \to H \nu\nu$ ($H \to hh \to 4 b$)
at $\sqrt{s} = 1.4$ TeV with $\mathcal{L} = 1500$ fb$^{-1}$ (orange) and $\sqrt{s} = 3$ TeV with $\mathcal{L} = 2000$ fb$^{-1}$ (blue). 
In both cases the solid line corresponds to a 70\% $b$-tagging efficiency and the dashed line to a 90\% $b$-tagging efficiency.
Shown for comparison are the LHC 95\% C.L. excluded region from present CMS 
$H \to hh \to 4 b$ searches~\cite{CMS:2017xxp} (red region) and the projected HL-LHC ($13$ TeV, $\mathcal{L} = 3$ ab$^{-1}$) expected 95\% C.L. 
exclusion sensitivity (dashed red line).}
\label{Sensitivity_H_hh_3000}
\end{center}

\end{figure}

In Figure~\ref{Sensitivity_H_hh_3000} we show the corresponding sensitivity of CLIC with $\sqrt{s} = 1.4$ TeV (blue) 
and $\sqrt{s} = 3$ TeV (orange) for 70\% $b$-tagging (solid) and 
90\% $b$-tagging (dashed) efficiencies, together with the present limits 
from CMS $H \to hh \to b\bar{b}b\bar{b}$ searches~\cite{CMS:2017xxp} with $\mathcal{L} = 35.9$ fb$^{-1}$ (solid red)
and the projected 95\% C.L. sensitivity for HL-LHC with $\mathcal{L} = 3$ ab$^{-1}$ (dashed red)
based on a $\sqrt{\mathcal{L}}$ scaling w.r.t.~to the present expected exclusion sensitivity from~\cite{CMS:2017xxp}.
As Figure~\ref{Sensitivity_H_hh_3000} highlights, CLIC would greatly surpass the sensitivity of HL-LHC
to resonant di-Higgs production: for a c.o.m. energy $\sqrt{s} = 1.4$ TeV the increase in sensitivity w.r.t. HL-LHC ranges from a factor $30 - 50$ for 
$m_2 \lesssim 400$ GeV, to roughly a factor $10$ for $m_2 \sim 1$ TeV. For $\sqrt{s} = 3$ TeV the increase in sensitivity is a factor $50$ or larger in the entire 
mass range $m_2 \in [250\,\mathrm{GeV}, \, 1\, \mathrm{TeV}]$, reaching two orders of magnitude sensitivity increase for $m_2 < 400$ GeV and $m_2 > 800$ GeV. 
At the same time, our results show that increasing the $b$-tagging efficiency above the 
70\% working point would benefit the reach of this search at CLIC substantially. In our work we specifically explore a 90\% working point, but a less 
extreme increase of the $b$-tagging efficiency would display a comparable associated sensitivity increase.  

\vspace{1mm}

Altogether, the results of this section show that resonant di-Higgs production searches are a prominent and very sensitive probe of 
heavier Higgs bosons with CLIC. In the remainder of this work, we explore the sensitivity of these searches to the 
existence of a new singlet-like scalar interacting with the 
SM Higgs, and the implications for the properties of the EW phase transition in the early Universe.

\section{Singlet scalar extension of the Standard Model}
\label{sec:xSM}

\vspace{-3mm}

The (real) singlet extension of the SM is a simple scenario that can capture the phenomenology of the Higgs sector in 
more complete theories beyond the SM (like the NMSSM and Twin Higgs). At the same time, 
it constitutes a paradigm for achieving a strongly first order EW phase transition that could generate the observed matter-antimatter 
asymmetry of the Universe.
The phenomenology of the SM extended by a real scalar singlet $S$ (SM + $S$) has been widely studied in the 
literature 
(see 
e.g.~\cite{Profumo:2007wc,Barger:2007im,Espinosa:2011ax,No:2013wsa,Profumo:2014opa,Chen:2014ask,
Robens:2015gla,Buttazzo:2015bka,Bojarski:2015kra,Kotwal:2016tex, Huang:2016cjm, Huang:2017jws,Chen:2017qcz,Lewis:2017dme}), 
including the connection to the EW phase 
transition~\cite{Profumo:2007wc,Espinosa:2011ax,No:2013wsa,Profumo:2014opa,Kotwal:2016tex,Huang:2016cjm,Chen:2017qcz} (see also~\cite{Curtin:2014jma,Chala:2016ykx}). 
We analyse here the sensitivity of CLIC to the parameter space leading to a first order EW phase transition by casting the 
results from the previous sections in terms of the SM + $S$ scenario. We also explore the complementarity of CLIC 
with other probes of the EW phase transition -- favoured parameter space in this scenario from HL-LHC 
and future colliders such as {\sl FCC-ee}~\cite{Huang:2016cjm,Chen:2017qcz}.

\subsection{Model and theoretical constraints}

We consider the most general form for the SM + $S$ scalar potential that depends on a Higgs 
doublet $\Phi$ and real singlet $S$ (see e.g.~\cite{Profumo:2007wc,Espinosa:2011ax}):
\begin{eqnarray}
\label{ScalarPotential1}
& V(\Phi,S) =\ds  -\mu^2 \left( \Phi^\dagger \Phi \right) + \lambda \left( \Phi^\dagger \Phi \right)^2 + \frac{a_1}{2} \left( \Phi^\dagger \Phi \right) S & \nonumber \\
& \ds + \frac{a_2}{2} \left( \Phi^\dagger \Phi \right) S^2 + b_1 S + \frac{b_2}{2} S^2 + \frac{b_3}{3} S^3 + \frac{b_4}{4} S^4 . &
\end{eqnarray}
Upon EW symmetry breaking, $\Phi \to (v+h) / \sqrt{2}$ with $v = 246$ GeV.
We note that a shift in the singlet field $S + \delta S$ does not lead to any change in the physics, which may be used to 
choose a vanishing vev for the singlet field in the EW broken minimum by requiring $b_1 = - a_1 v^2/4$. This is the choice we adopt in the following.
Once the EW symmetry is broken, the singlet $S$ and the SM Higgs $h$ mix in the presence of $a_1$, 
yielding two mass eigestates $h_1$, $h_2$. We identify $h_1$ with the $125$ GeV Higgs boson, and $h_2$ with the heavy state $H$ discussed in the previous sections.
The masses $m_1 = 125$ GeV, $m_2$ and the singlet-doublet mixing angle $\theta$ are related to the scalar potential parameters as
\begin{eqnarray}
\label{ScalarPotential_vs_masses}
a_1 &=& \frac{m_1^2 - m_2^2}{v}\, 2 \,{\rm sin}\,\theta \,{\rm cos}\,\theta \nonumber \\
b_2 + \frac{a_2 \, v^2}{2} & = & m_1^2 \,{\rm sin}^2\theta + m_2^2\, {\rm cos}^2\theta \\
\lambda & = & \frac{m_1^2 \,{\rm cos}^2\theta + m_2^2 \,{\rm sin}^2\theta}{2 \, v^2} \nonumber
\end{eqnarray}
with $\mu^2 = \lambda\, v^2$. In the following we consider as independent parameters for our analysis the set 
$\left\lbrace v,\, m_1,\,m_2,\,\theta,\,a_2,\,b_3,\,b_4 \right\rbrace$.

\vspace{2mm}

In order to obtain a viable SM + $S$ scenario, we need to satisfy several theoretical constraints which we discuss below:
\vspace{2mm}

\noindent $\bullet$ {\sl (Perturbative) unitarity and perturbativity}: The size of the quartic scalar couplings in eq.~\eqref{ScalarPotential1} is constrained 
by perturbative unitarity of the partial wave expansion of scattering amplitudes. The bound $\left|a_0 \right| \leq 0.5$ for the leading order 
term in the partial wave expansion of the $h_2 h_2 \to h_2 h_2$ scattering amplitude, $a_0 (h_2 h_2 \to h_2 h_2) = 3 b_4/(8\pi)$, yields
$b_4 < 4\pi/3$ (see e.g.~\cite{Lewis:2017dme}). In addition, we require perturbative values for 
$a_2$ and $b_3/v$: $\left|a_2 \right| < 4\pi$, $\left|b_3 \right|/v < 4\pi$.

\vspace{2mm}

\noindent $\bullet$ {\sl Boundedness from below of scalar potential}: We require the absence of runaway directions in the scalar potential~\eqref{ScalarPotential1} 
at large field values.  Along the $h$ and $S$ directions, this leads respectively to the bounds $\lambda > 0$ and $b_4 > 0$. 
For $a_2 < 0$ we further require $a_2 > - 2 \sqrt{\lambda\, b_4}$ to ensure boundedness from below along an arbitrary field direction.

\vspace{2mm}

\noindent $\bullet$ {\sl Absolute stability of EW vacuum}: First, the EW vacuum $\left( \left\langle h \right\rangle, \left\langle S \right\rangle \right) = (v, 0)$ 
must be a minimum. On one hand, this requires $b_2 > 0$, which by virtue of~\eqref{ScalarPotential_vs_masses} yields an upper bound on the value of $a_2$
\begin{equation}
a_2 < \frac{2}{v^2} (m_1^2 \,{\rm sin}^2\theta + m_2^2\, {\rm cos}^2\theta) \, .
\end{equation}
On the other hand, for $(v, 0)$ to be a minimum the determinant of the scalar squared-mass matrix has to be positive
\begin{equation}
\mathrm{Det} \left. \left(\begin{array}{c c}
              \partial^2 V / \partial h^2  & \partial^2 V / \partial h \partial S \\
              \partial^2 V / \partial h \partial S & \partial^2 V / \partial S^2
             \end{array}\right)\right|_{(v,0)} \equiv \mathrm{Det} \mathcal{M}^2_S = 2\,\lambda v^2\, b_2 - \frac{a_1^2\,v^2}{4} > 0 \, .
\end{equation}
In addition, we require that the EW vacuum is the absolute minimum of the potential. The conditions for this are discussed in 
detail in~\cite{Espinosa:2011ax}, and we summarise them here. It will prove convenient to define the quantities
\begin{eqnarray}
\label{JR_parametrization}
\overline{\lambda}^2 &  \equiv & \lambda\, b_4 - \frac{a_2^2}{4} \nonumber \\
m_{*} & \equiv & \frac{\lambda\, b_3}{3} - \frac{a_2\, a_1}{8} \\
\mathcal{D}^2(S) & \equiv & v^2 \left( 1 - \frac{a_1\,S}{2\lambda v^2} - \frac{a_2\,S^2}{2\lambda v^2} \right) 
\end{eqnarray}
with $h^2 = \mathcal{D}^2(S)$ corresponding to the minimization condition $\partial V/\partial h = 0$ for values $h \neq 0$. 
From the analysis of~\cite{Espinosa:2011ax}, we immediately find that a sufficient (though not necessary) condition for 
the EW vacuum to be the absolute minimum of $V$ is given by
\begin{equation}
\label{AbsEWMin_1}
\overline{\lambda}^2 > \frac{m_{*}^2 \, v^2}{16 \,\mathrm{Det} \mathcal{M}^2_S} \, .
\end{equation}
When~\eqref{AbsEWMin_1} is not satisfied, there exists for $\overline{\lambda}^2 > 0$ 
a minimum $S = \omega$ along $\mathcal{D}^2(S)$ which is deeper than the EW vacuum, 
and in order for the EW vacuum to still be the absolute minimum of $V$, it is necessary that $\mathcal{D}^2( \omega) < 0$ (in order for this new minimum 
to be unphysical). In addition, in this case we also need to require that no new minimum exists along the $h = 0$ field direction which is deeper than 
the EW one. The extrema along this direction are given by the real solutions of the equation
\begin{equation}
\label{AbsEWMin_2}
b_4 S^3 + b_3 S^2 + b_2 S + b_1 = 0 \, .
\end{equation}
Finally, when $\overline{\lambda}^2 < 0$ a necessary and sufficient condition for the EW vacuum to be the absolute minimum of $V$ is the absence of a deeper minimum 
along the $h = 0$ field direction, which we have just discussed above.

\begin{figure}[h]

\begin{center}
\includegraphics[width=0.99\textwidth]{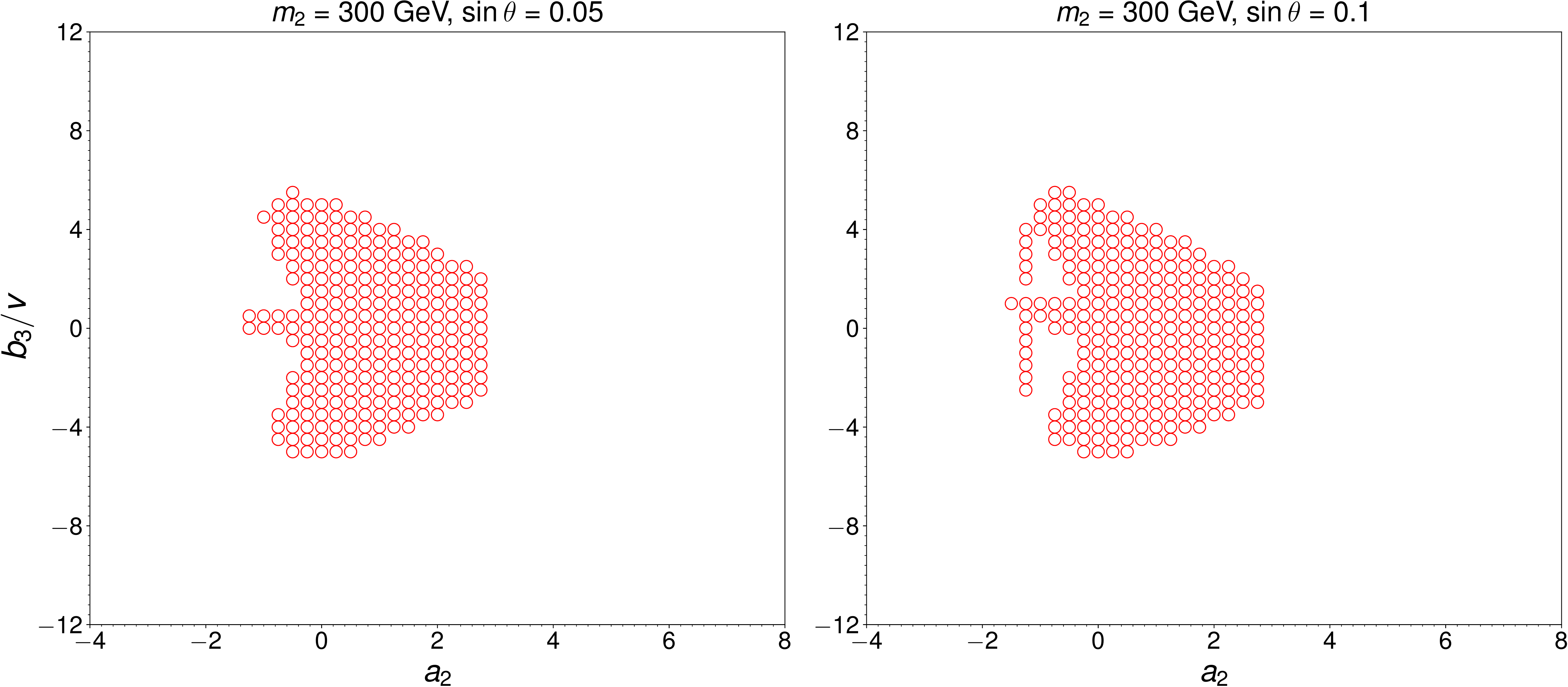}

\vspace{-2mm}

\caption{\small Region of parameter space in ($a_2$, $b_3/v$) and fixed $m_2 = 300$ GeV and ${\rm sin}\,\theta = 0.05$ (left), ${\rm sin}\,\theta = 0.1$ 
(right), compatible with the requirements of unitary, perturbativity and absolute stability of the EW vacuum. The parameter $b_4$ has been scanned over 
(see text for details).}
\label{EWVAcuum_Plot_300}
\end{center}

\end{figure}
\begin{figure}[h]

\begin{center}
\includegraphics[width=0.99\textwidth]{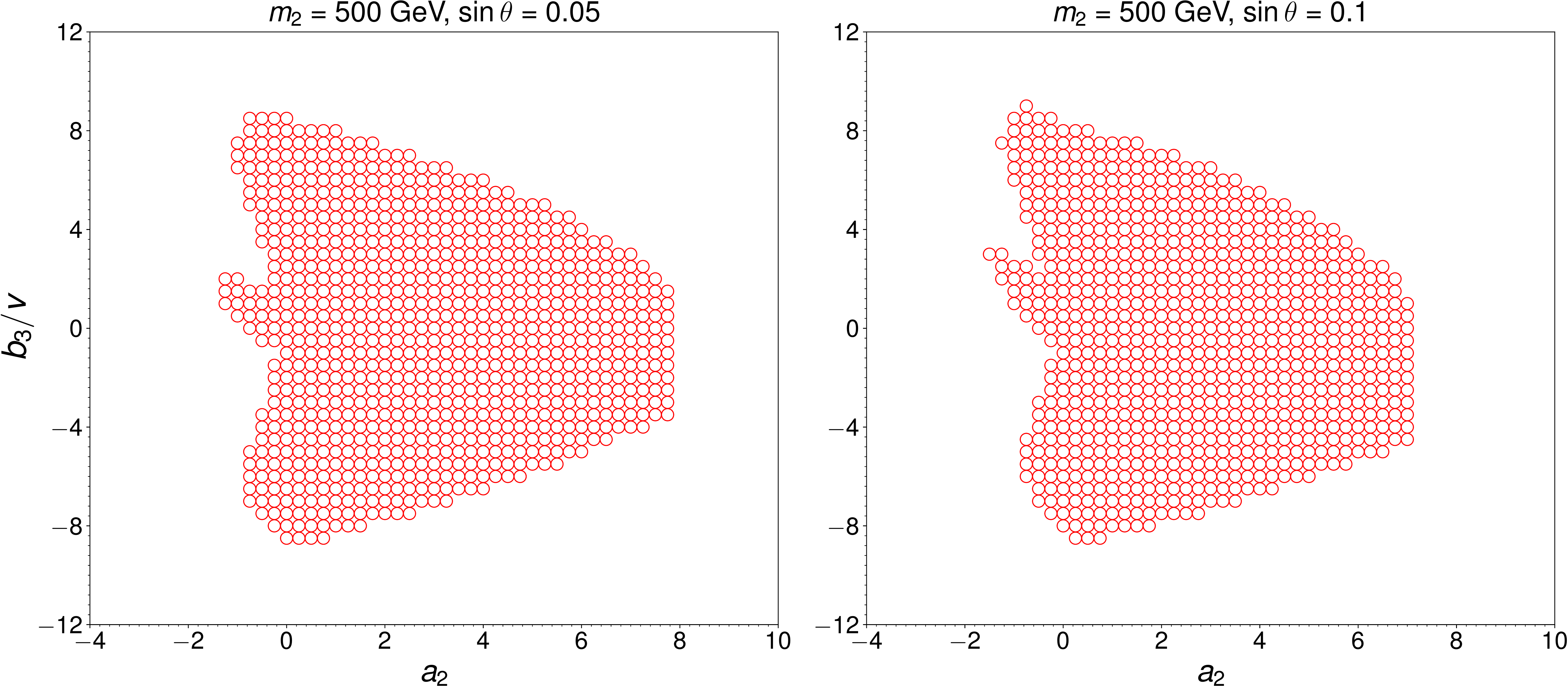}

\vspace{-4mm}

\caption{\small Same as Figure~\ref{EWVAcuum_Plot_300}, but for $m_2 = 500$ GeV.}
\label{EWVAcuum_Plot_500}
\end{center}

\end{figure}
\begin{figure}[t]

\begin{center}
\includegraphics[width=0.99\textwidth]{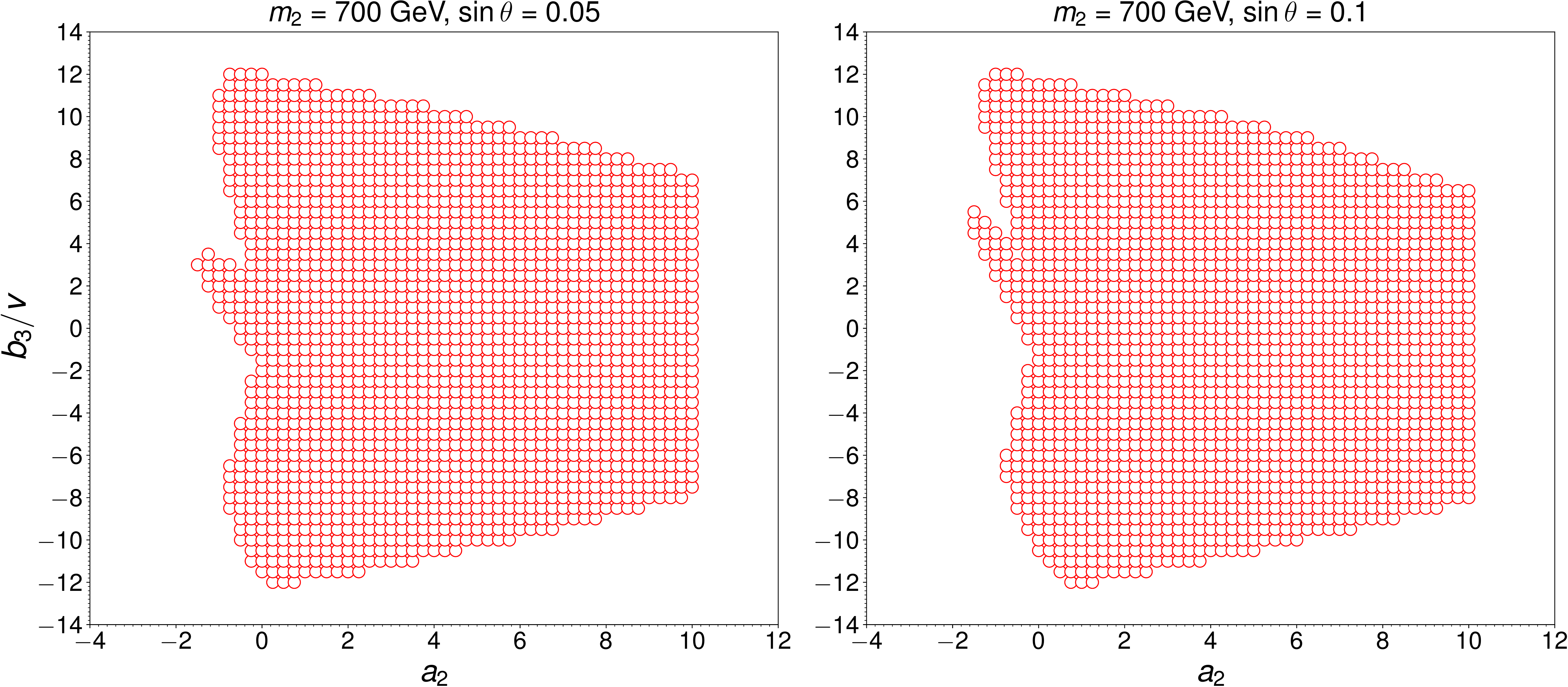}

\vspace{-2mm}

\caption{\small Same as Figure~\ref{EWVAcuum_Plot_300}, but for $m_2 = 700$ GeV.}
\label{EWVAcuum_Plot_700}
\end{center}

\end{figure}

In Figures~\ref{EWVAcuum_Plot_300}--\ref{EWVAcuum_Plot_700},  we show, for fixed values of $m_2 = 300$ GeV, $500$ GeV, $700$ GeV 
and ${\rm sin}\,\theta = 0.1,\,0.05$, the points that satisfy the above 
requirements in the plane $a_2$, $b_3/v$, with the parameter $b_4$ being scanned over. 
We find that, for a given choice of $(a_2$, $b_3/v)$, the requirements are generically satisfied more robustly as $b_4$ 
increases\footnote{This is true except in certain regions of $a_2 < 0$, where ``islands of stability" in the parameter $b_4$ exist (that is, a very narrow range of 
$b_4$ within $[0 ,\, 4\pi/3]$ where the EW vacuum is the absolute minimum of the potential. These regions are however not relevant for the subsequent 
EW phase transition discussion, and we disregard them in the following.}, and as such we demand that there is a value of $b_4 \in [0 ,\, 4\pi/3]$ above which 
the EW vacuum is the absolute minimum of the potential. 

\vspace{-3mm}

\vspace{-3mm}

Before moving on to the next section, we note that for large values of $a_2$ and $b_3$ the 1-loop corrections may become 
important and might allow for new regions that fulfill the above 
stability/unitarity/perturbativity conditions (see the discussion in~\cite{Chen:2017qcz}), particularly for low values of $m_2$, 
for which such regions with large $a_2$ and/or $b_3$ do no satisfy these requirements 
at tree-level (see Figures~\ref{EWVAcuum_Plot_300}-\ref{EWVAcuum_Plot_700}).
We leave an investigation of the impact of 1-loop corrections on the above theoretical constraints for the future.
We also note that, as compared to~\cite{Chen:2017qcz}, our analysis has a smaller range of allowed values for 
$b_4$ which is partially responsible (together with the different chosen range for $m_2$) for the different shape of the tree-level allowed region. 

\subsection{EW phase transition in the SM + $S$}

The EW symmetry is (generally) restored at high temperatures $T \gg v$. 
EW symmetry breaking then occurs when the temperature of the Universe drops due to expansion, and it 
becomes energetically favorable for the Higgs field $\Phi$ to acquire a non-zero expectation value
$\varphi_h = v_T \neq 0$. When there exists a potential barrier separating the symmetric vacuum $\varphi_h = 0$ from the broken 
one $v_T$, the EW phase transition is of first order. The temperature at which the two vacua become degenerate in energy 
is known as the critical temperature $T_c$, and the EW phase transition is considered to be “strongly first order” 
if\footnote{A more accurate criterion can be obtained by considering the ``nucleation" temperature $T_n$ at which the phase transition actually takes place, 
and requiring $v_{T}(T_n)/T_n \gtrsim 1$. It is nevertheless a reasonable approximation in general to consider $v_{T}(T_c)/T_c \gtrsim 1$ instead.} 
$v_{T}(T_c)/T_c \gtrsim 1$.

For the analysis of the EW phase transition in the SM + $S$ scenario, we adopt in the following a 
conservative strategy: It is known that including the 1-loop $T = 0$ (Coleman-Weinberg) contributions 
to the effective potential introduces a gauge-dependence\footnote{This gauge-dependence arises from 
the Goldstone and gauge boson contribution to the Coleman-Weinberg potential, as well as to the cubic term of the finite-temperature potential 
in the high-$T$ expansion (see~\cite{Patel:2011th} for a detailed discussion).} in the evaluation of various phase transition parameters, such as
$T_c$~\cite{Dolan:1973qd,Nielsen:1975fs,Patel:2011th}. However for a singlet-driven first order EW phase transition as in the SM + $S$, 
the properties of the transition are dominantly determined by tree-level effects. It is then possible in a first approximation 
to perform the analysis of the phase transition 
using the tree-level 
potential~\eqref{ScalarPotential1} augmented by the $T^2$ terms from the high-$T$ expansion of the finite-temperature effective potential (see 
e.g.~\cite{Espinosa:2011ax}):
\begin{equation}
\label{T2_terms}
V_{T^2} = \left(\frac{c_h}{2}h^2 + \frac{c_s}{2}S^2 + c_t S \right) T^2\, ,  
\end{equation}
where
\begin{eqnarray}
c_h &=& \frac{1}{48} \left(9 g^2 + 3 g'^{2} + 12 y_t^2 + 24 \lambda + 2 a_2   \right)\nonumber \\
c_s &=& \frac{1}{12} \left( 2 a_2 + 3 b_4  \right) \nonumber \\
c_t &=& \frac{1}{12} \left( a_1 + b_3  \right) \nonumber
\end{eqnarray}
as these are manifestly gauge invariant\footnote{The last term in~\eqref{T2_terms} is gauge invariant at 1-loop, but not necessarily at 
higher loop order~\cite{Chen:2017qcz,Profumo:2014opa}. Still, we choose here to keep it in the analysis (in contrast 
to~\cite{Chen:2017qcz,Profumo:2014opa}, where such term is discarded).}.
This approach, which we take in the present work, 
nevertheless disregards 1-loop terms that could be numerically important in certain regions of parameter space, particularly for large values of 
$a_2$ and/or $b_3$, strengthening the phase transition in those regions. We believe the choice made here then provides a conservative 
prediction for a strongly first order EW phase transition.

\vspace{2mm}

In the following we use the numerical programme {\sc CosmoTransitions}~\cite{Wainwright:2011kj} (v2.0.2) to find the points in parameter space with a viable 
strongly first order EW phase transition, for fixed values of $m_2$ and ${\rm sin}\,\theta$ while scanning over $a_2$, $b_3$ and $b_4$. Specifically, for each scan point we 
evolve the effective potential (combining~\eqref{ScalarPotential1} and~\eqref{T2_terms}) from $T = 0$ up and look for coexisting and degenerate 
phases at some temperature(s) $T^*_i = T_c$. We consider the point to have a strongly first order EW phase transition when at (any) such temperature there is coexistence of 
a phase with $\varphi_h = 0$ (irrespectively of the singlet vacuum expectation value) and a phase with $\varphi_h  = v_T$, separated by a potential barrier and such that 
$v_T/T_c > 1$.
The results of our EW phase transition scan are shown in Figures~\ref{EWPT_Plot_300}--\ref{EWPT_Plot_700}, 
with the same stability/unitarity/perturbativity requirements as in 
Figures~\ref{EWVAcuum_Plot_300}--\ref{EWVAcuum_Plot_700}.~We also overlay the projected sensitivities from CLIC, as well as those from 
HL-LHC and {\sl FCC-ee}, all discussed in the next section.
%
Our EW phase transition scan shows that, as the mass $m_2$ increases, the values of $a_2$ and $b_3/v$ required to achieve a strongly first order transition 
also increase substantially, approaching the perturbativity limit (particularly for $a_2$) for $m_2 \sim 700 - 800$ GeV. This yields a clear target reach for 
high-energy colliders regarding a singlet-driven EW phase 
transition\footnote{We emphasize again that the 1-loop Coleman-Weinberg and finite-$T$ terms of the effective potential 
disregarded here will have some impact on the precise shape of the parameter space region yielding a strongly first order EW phase transition, 
and the value of $m_2$ above which such a strong transition stops being feasible. Yet, the bound $m_2 \lesssim 700 - 800$ GeV
will not be significantly modified.}.

\subsection{CLIC sensitivity to the SM + $S$: probing the EW phase transition}

We analyse here the CLIC prospects for probing the parameter space leading to a strongly first order EW phase transition 
in the SM + $S$ scenario, based on the results from the previous sections. In addition, we discuss the   
complementarity with probes of this parameter space via other possible future colliders, 
such as {\sl FCC-ee}~\cite{Huang:2016cjm}, as well as from the HL-LHC.

\begin{figure}[h]

\begin{center}
\includegraphics[width=0.99\textwidth]{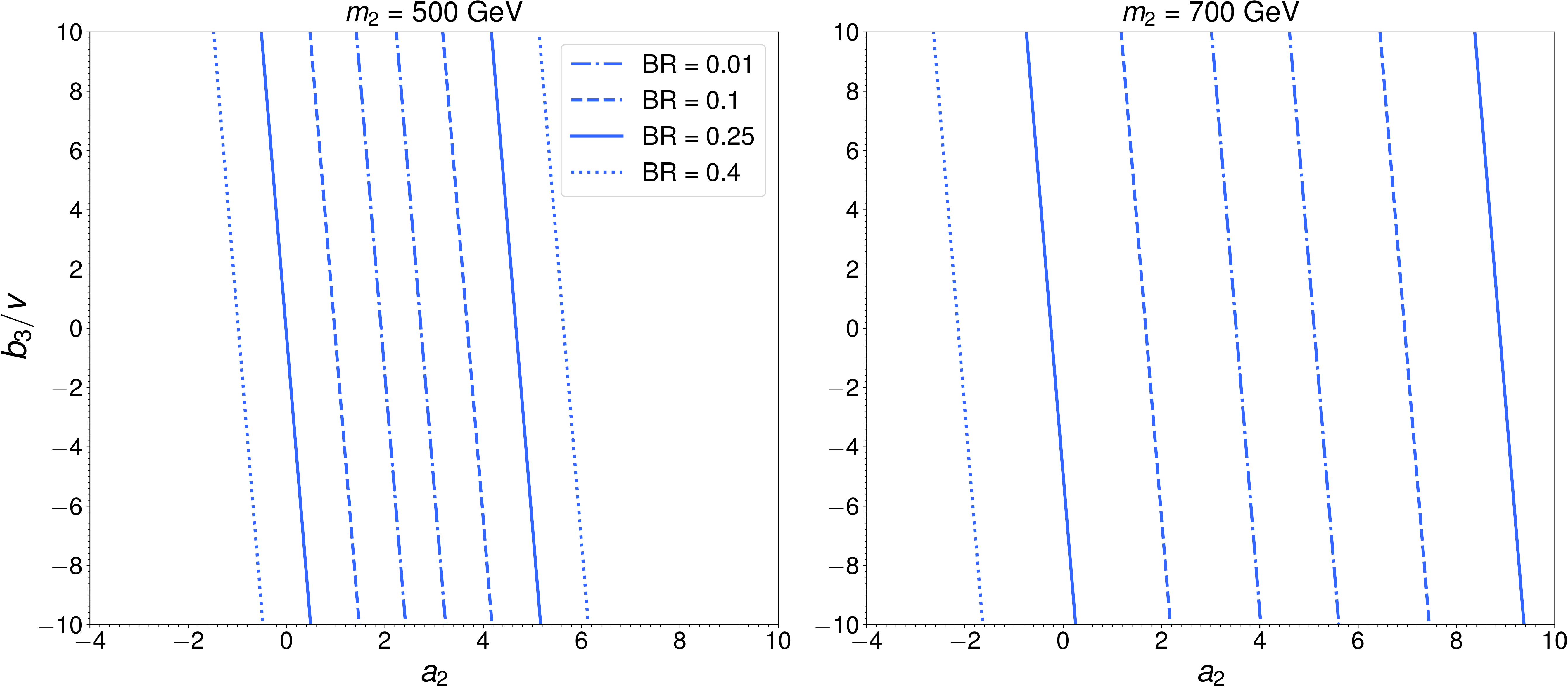}

\vspace{-2mm}

\caption{\small Branching fraction $\mathrm{BR}(h_2\to h_1 h_1)$ values in the SM + $S$ scenario
for $\mathrm{sin}\, \theta = 0.05$ and $m_2 = 500$ GeV (left), $m_2 = 700$ GeV (right) 
in the plane ($a_2$, $b_3/v$).}
\label{Plot_hh_BR}
\end{center}

\end{figure}

Let us start by pointing out that due to the singlet-doublet mixing, the couplings of $h_1$ ($h_2$) 
to SM gauge bosons and fermions are universally rescaled w.r.t. the corresponding SM Higgs coupling values by 
$\mathrm{cos}\,\theta$ ($\mathrm{sin}\,\theta$). In addition to these, 
the tri-scalar interactions play an important role in the discussion of both di-Higgs production at colliders and the nature of the EW
phase transition. Specifically, we focus on the interactions  
$\lambda_{211}h_2\,h_1\,h_1$ and $\lambda_{111}\,h_1\,h_1\,h_1$, which follow from \eqref{ScalarPotential1} after EWSB, with
\bea
\label{g211}
\lambda_{211} &=& \frac{1}{4}\left[ a_1 \, c_{\theta}^3 + 4 v (a_2 -3 \lambda)\, c_{\theta}^2 s_{\theta} 
 - 2 (a_1 -2\,b_3)\, c_{\theta} s^2_{\theta} -2a_2 v \, s^3_{\theta} \right] \nonumber \\
\lambda_{111} &=& \lambda v \, c_{\theta}^3 + \frac{1}{4} a_1 \, c_{\theta}^2 s_{\theta} +
\frac{1}{2} a_2\,v\, c_{\theta} s^2_{\theta} + \frac{b_3}{3} s^3_{\theta} 
\eea
with $c_{\theta} \equiv \mathrm{cos} \theta$ and $s_{\theta} \equiv \mathrm{sin} \theta$.
The coupling $\lambda_{211}$ controls the partial width of the decay $h_2 \to h_1 h_1$ for $m_2 > 250$ GeV, given by 
\begin{eqnarray}
\Gamma_{h_2 \to h_1 h_1} = \frac{\lambda_{211}^2 \sqrt{1 - \ds 4\, m_1^2 / m_2^2 } }{8 \pi m_2}\, . 
\label{partialWidthh1h1}
\end{eqnarray}
Denoting by $\Gamma^{\mathrm{SM}}(m_2)$ the total width of a SM-like Higgs with mass $m_2$ (as given {\it e.g.} in~\cite{Heinemeyer:2013tqa}), 
the branching fraction $\mathrm{BR}(h_2\to h_1 h_1)$ is simply given by 
\begin{eqnarray}
\mathrm{BR}(h_2\to h_1 h_1) = \frac{\Gamma_{h_2 \to h_1 h_1}}{\mathrm{sin}^2 \theta \; \Gamma^{\mathrm{SM}}(m_2) + \Gamma_{h_2 \to h_1 h_1}} \, .
\end{eqnarray}
In the limit of high $m_2$ masses, this branching fraction is expected to be fixed by the Equivalence Theorem\footnote{We are indebted to  
Andrea Tesi for reminding us of this.}, $\mathrm{BR}(h_2\to h_1 h_1) \simeq 0.25$, but different values of $a_2$ and $b_3$ can lead to some 
departure from this expectation. We show in Figure~\ref{Plot_hh_BR} the values of $\mathrm{BR}(h_2\to h_1 h_1)$ for 
$m_2 = 500,\, 700$ GeV and $\mathrm{sin}\, \theta = 0.05$ for illustration.
At the same time, the production cross section for $h_2$ normalized to the SM value (for a given mass $m_2$) takes in the case of the 
SM + $S$ scenario the very simple form $\sigma_S/\sigma^{\mathrm{SM}}_S = \mathrm{sin}^2 \theta$, due to the universal rescaling discussed above. 

\begin{figure}[h]

\begin{center}
\includegraphics[width=0.99\textwidth]{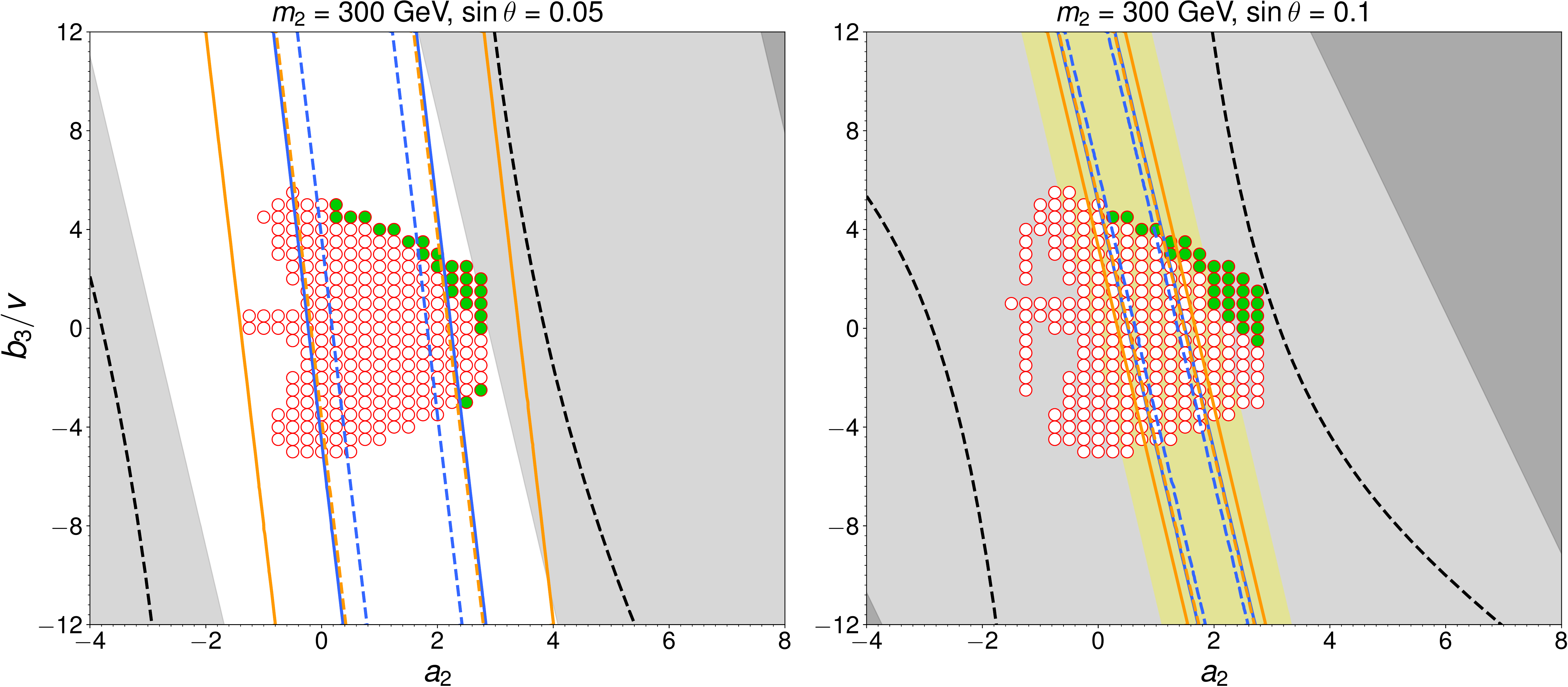}

\vspace{-2mm}

\caption{\small Region of parameter space in ($a_2$, $b_3/v$) for $m_2 = 300$ GeV and ${\rm sin}\,\theta = 0.05$ (left), ${\rm sin}\,\theta = 0.1$ 
(right) within the 95\% C.L. sensitivity reach of resonant di-Higgs production 
searches at CLIC with $\sqrt{s} = 1.4$ TeV (orange) and $\sqrt{s} = 3$ TeV (blue) for a $b$-tagging efficiency of 
$70$\% (solid) and $90$\% (dashed): CLIC sensitivity region is that not contained within each pair of (sensitivity) lines. 
Overlaid are the SM + $S$ points compatible with unitary, perturbativity and absolute stability of the EW vacuum from Figure~\ref{EWVAcuum_Plot_300}, 
and those yielding a strongly first order EW phase transition (green points). The dashed black lines correspond to the CLIC sensitivity to 
Higgs self-coupling deviations w.r.t. the SM $\delta \lambda_{111} = 0.19$. The yellow region (only for ${\rm sin}\,\theta = 0.1$) corresponds to the 
projected sensitivity of $p p \to h_2 \to Z Z$ searches at HL-LHC. The region within reach of a measurement of $\delta \sigma_{Zh}$ at CLIC ({\sl FCC-ee}) is shown in 
dark (light) grey.}
\label{EWPT_Plot_300}
\end{center}

\end{figure}

\begin{figure}[h]

\begin{center}
\includegraphics[width=0.99\textwidth]{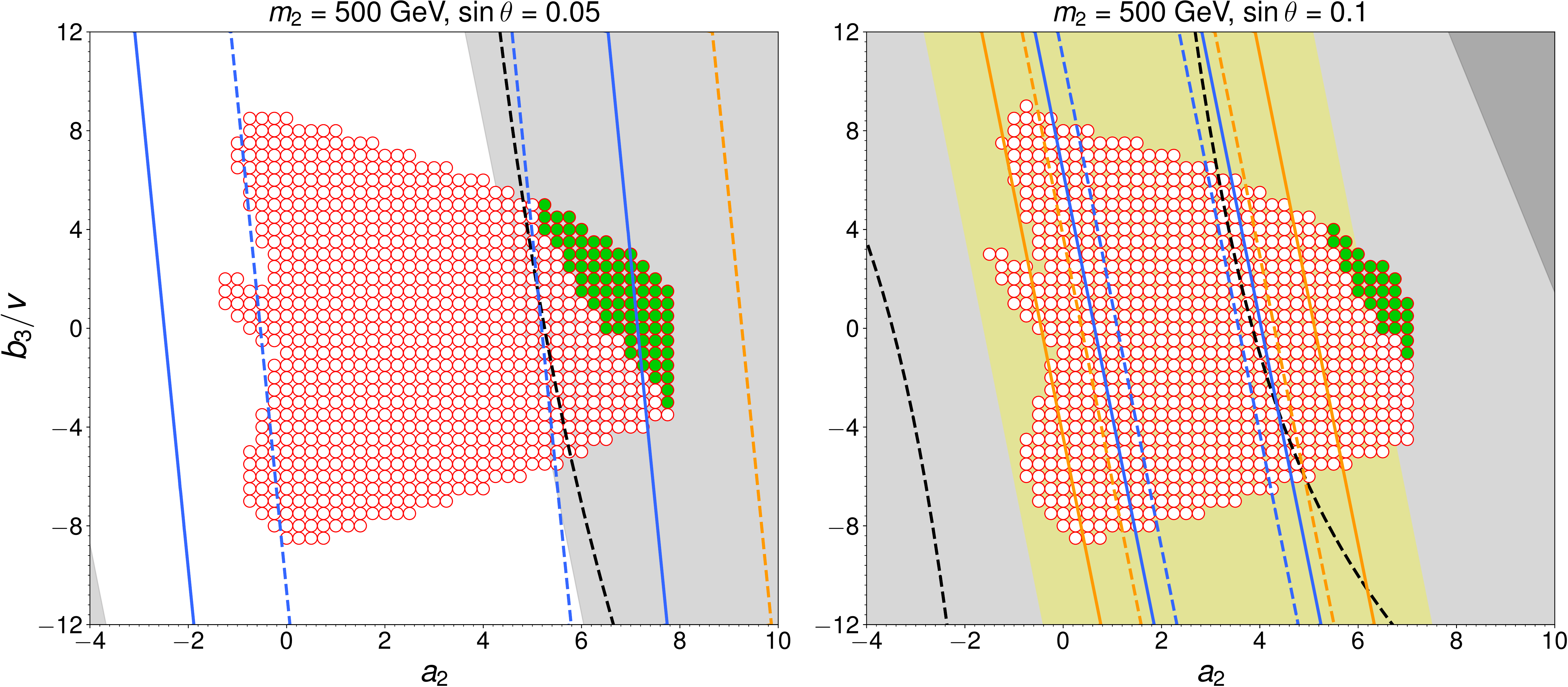}

\vspace{-2mm}

\caption{\small Same as Figure~\ref{EWPT_Plot_300}, but for $m_2 = 500$ GeV.}
\label{EWPT_Plot_500}
\end{center}

\end{figure}

\begin{figure}[h]

\begin{center}
\includegraphics[width=0.99\textwidth]{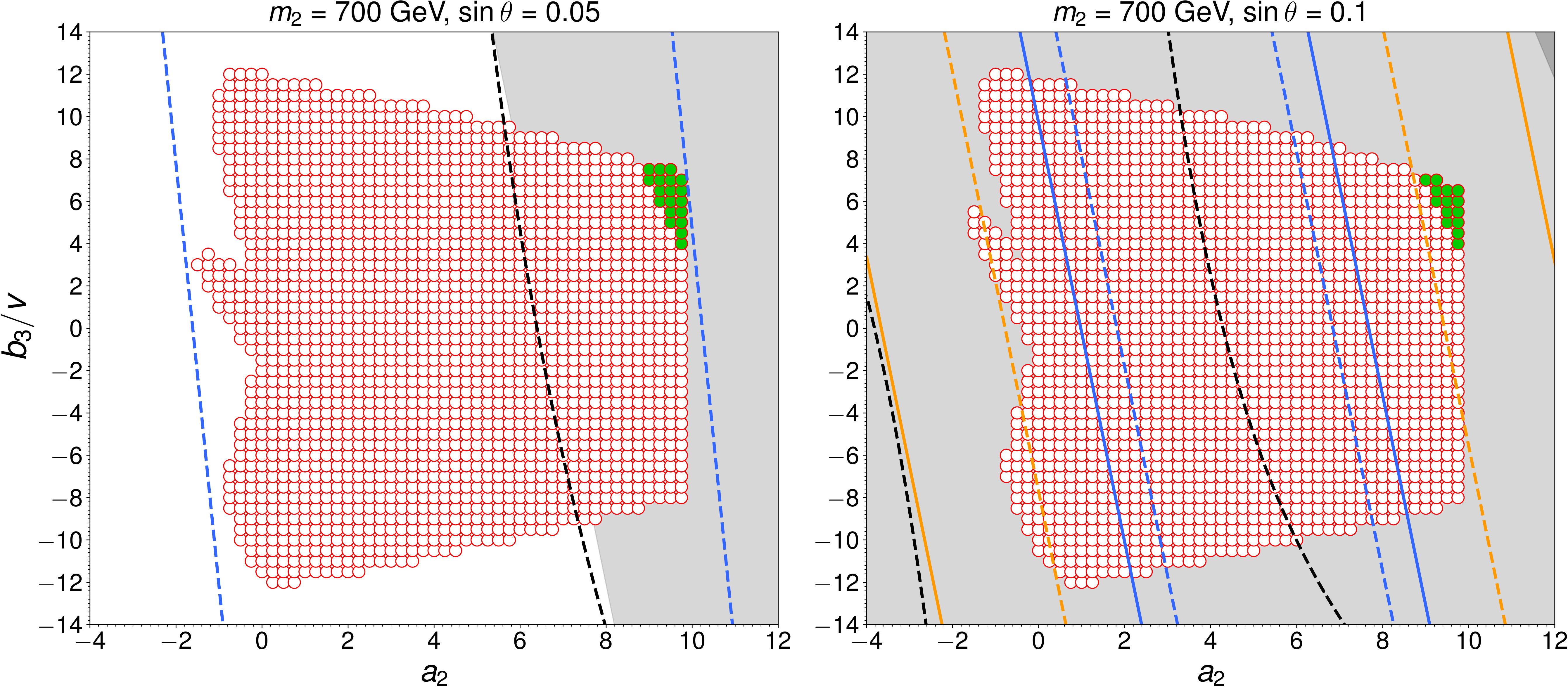}

\vspace{-2mm}

\caption{\small Same as Figure~\ref{EWPT_Plot_300}, but for $m_2 = 700$ GeV.}
\label{EWPT_Plot_700}
\end{center}

\end{figure}

With all these ingredients, we can readily interpret both the HL-LHC and CLIC sensitivities to the parameter space of the 
SM + $S$ scenario. First, we note that the projected HL-LHC sensitivity to the singlet-doublet mixing from a global fit to the measured
125 GeV Higgs signal strengths is~\cite{ATL-PHYS-PUB-2014-016} $\mathrm{sin}\, \theta \simeq 0.18$ (assuming negligible theory uncertainties; 
taking into account the present theory 
uncertainties the projected value is $\mathrm{sin}\, \theta \simeq 0.25$). In the present work we have thus always
considered $\mathrm{sin}\, \theta $ to be smaller than this value.  
The interpretation of the sensitivity of direct searches in CLIC
(discussed in sections~\ref{subsec:VV} and~\ref{subsec:hh}) in the context of the 
SM + $S$ scenario is shown in Figures~\ref{EWPT_Plot_300}--\ref{EWPT_Plot_700} for $m_2 = 300$, $500$, $700$ GeV and $\mathrm{sin}\, \theta = 0.1, \,0.05$: we 
show the resonant di-Higgs production 
sensitivity of CLIC with $\sqrt{s} = 1.4$ TeV (orange) and $\sqrt{s} = 3$ TeV (blue) for a respective $b$-tagging efficiency of 
$70$\% (solid) and $90$\% (dashed), with CLIC able to probe the region not contained within each pair of sensitivity lines. 
For the case $\mathrm{sin}\, \theta = 0.1$ (for $\mathrm{sin}\, \theta = 0.05$ there is no sensitivity) we also show the HL-LHC sensitivity to the process 
$p p \to h_2 \to Z Z$ (see section~\ref{subsec:VV}) as a shadowed yellow region.

In addition to the direct searches for $h_2$, we consider here two indirect collider probes of the SM + $S$ scenario: 

\noindent {\sl(i)} The measurement of the 125 GeV Higgs self-coupling $\lambda_{111}$. 
The projected sensitivity to the Higgs self-coupling at CLIC, combining the $\sqrt{1.4}$ TeV and $\sqrt{3}$ TeV runs 
is $\delta \lambda_{111} \equiv \left|\lambda^{\mathrm{SM}+S}_{111} - \lambda^{\mathrm{SM}}_{111}\right|/\lambda^{\mathrm{SM}}_{111} = 19\%$
(for a choice of beam polarization similar to the one considered in this work)~\cite{Abramowicz:2016zbo}, 
with $\lambda^{\mathrm{SM}}_{111} = \lambda\,v = 31.8$ GeV being the self-coupling value in the SM. 
For the Higgs self-coupling in the SM + $S$ scenario, we consider both the tree-level contribution from~\eqref{g211} and the 1-loop contribution
computed to order $\mathrm{sin}\, \theta $ and given by~\cite{Chen:2017qcz} (note the different $\lambda_{111}$ normalization in our work w.r.t.~\cite{Chen:2017qcz}):
\be
\Delta \lambda^{\mathrm{1-loop}}_{111} = 
\frac{1}{16\pi^2} \left(\frac{a_2^3 \,v^3}{12\, m_2^2}  
+ \frac{a_2^2 \,b_3\,v^2}{2\,m_2^2}\,\mathrm{sin}\, \theta \right)\, .
\label{g111_loop}
\ee
We then consider the region accesible to CLIC 
as $\left|(\lambda_{111} + \Delta \lambda^{\mathrm{1-loop}}_{111}) - \lambda^{\mathrm{SM}}_{111}\right|/\lambda^{\mathrm{SM}}_{111} = 0.19$ 
(the tree-level and 1-loop contributions given respectively by~\eqref{g211} and~\eqref{g111_loop}), depicted in 
Figures~\ref{EWPT_Plot_300}--\ref{EWPT_Plot_700} as a dashed-black curve.
We nevertheless stress that it is not at all clear that the information on $\lambda^{\mathrm{SM}+S}_{111}$ from the non-resonant di-Higgs signal 
can be extracted from the data independently from the resonant di-Higgs contribution.  
In particular, since the non-resonant Higgs pair invariant mass distribution $m_{hh}$ peaks around $300 - 400$ GeV  
(see~\cite{LCD-Note-2012-014}), for masses $m_2 \lesssim 500$ GeV disentangling the two contributions might be challenging.

\vspace{1mm}

\noindent {\sl(ii)} The measurement of the Higgs associated production cross section $\sigma_{Zh}$ at CLIC and {\sl FCC-ee}. 
At CLIC, the expected precision in the determination of the associated production cross section for the 125 GeV Higgs is 
$\delta\sigma_{Zh} \equiv \left|\sigma_{Zh} - \sigma_{Zh}^{\mathrm{SM}}\right|/\sigma_{Zh}^{\mathrm{SM}} = 1.65\%$~\cite{Abramowicz:2016zbo}. 
A future circular $e^+ e^-$ collider like {\sl FCC-ee}
could reach a precision $\delta\sigma_{Zh} = 0.4\%$~\cite{Gomez-Ceballos:2013zzn,dEnterria:2016fpc}. For a small singlet-doublet mixing (as we are considering here), 
the deviation in the Higgs associated production cross section w.r.t. its SM value is approximately given 
by (see e.g.~\cite{Chen:2017qcz,Huang:2016cjm,Craig:2013xia}):
\be
\delta\sigma_{Zh} = \left| - \mathrm{sin}^2\, \theta + \frac{\lambda_{221}^2}{16\,\pi^2\,m_1^2} (1 - F(\tau)) \right|\, ,
\label{deltaZh}
\ee
where the first term is just the tree-level deviation and the second term corresponds to the leading 1-loop correction, 
with $\tau = m_1^2/(4 m_2^2)$ and $F(\tau)$, $\lambda_{221}$ given by
\be
F(\tau) = \frac{\mathrm{Arcsin}(\sqrt{\tau})}{\sqrt{\tau (1- \tau)}}\, ,
\label{loopZh}
\ee
\be
\lambda_{221} =  \frac{1}{2} \,a_2\,v \, c_{\theta}^3 + (b_3 - \frac{a_1}{2}) \, c_{\theta}^2 s_{\theta} + v (3 \lambda - a_2) \, c_{\theta} s^2_{\theta}
 + \frac{a_1}{4} \, s^3_{\theta} \, .
\label{g221_tree}
\ee

In Figures~\ref{EWPT_Plot_300}--\ref{EWPT_Plot_700} we show the indirect reach in the ($a_2,\, b_3/v$) plane for fixed $m_2$ and ${\rm sin}\,\theta$
through the measurement of $\delta \sigma_{Zh}$ both for CLIC (dark grey) and {\sl FCC-ee} (light grey).
For ${\rm sin}\,\theta = 0.1$, such a measurement of $\delta \sigma_{Zh}$ at {\sl FCC-ee} would yield the most powerful constraint on the SM + $S$ scenario,
allowing to access the entire parameter space of the model.
In contrast, for ${\rm sin}\,\theta = 0.05$ this measurement would yield a comparable sensitivity to that of the Higgs self-coupling, and would be less sensitive 
than resonant di-Higgs searches at CLIC for masses $m_2 \lesssim 500$ GeV.

\vspace{2mm}

The results from Figures~\ref{EWPT_Plot_300}--\ref{EWPT_Plot_700} also highlight that it would be possible in many cases to simultaneously access
via direct and indirect collider probes the region of parameter space yielding a strongly first order EW phase transition in the SM + $S$ scenario. 
This would allow to correlate the information from the various probes towards providing a robust test of the nature of the EW phase transition.

Before concluding, we emphasize that for a vanishing singlet-doublet mixing ${\rm sin}\,\theta \to 0$ 
(as is e.g.~the case in the $\mathbb{Z}_2$ symmetric limit of the SM + $S$ scenario)
the resonant di-Higgs signature also vanishes, while the indirect probes $\delta \lambda_{111}$ and $\delta \sigma_{Zh}$ have their sensitivity significantly reduced 
 (as deviations w.r.t.~the SM only occur at 1-loop via 
the parameter $a_2$), particularly for low masses $m_2$. Yet in this limit 
a strongly first order EW phase transition is still possible~\cite{Curtin:2014jma,Huang:2016cjm,Chen:2017qcz}.
The dominant probe of this parameter space region of the SM + $S$ (the so-called ``nightmare-scenario" for EW baryogenesis~\cite{Curtin:2014jma}) 
could be given by pair production of the singlet-like state $h_2$~\cite{Chen:2017qcz} (except for 
the case of exact $\mathbb{Z}_2$ symmetry, $h_2$ would decay into SM states), and we note that a high-energy $e^+ e^-$ collider 
like CLIC could provide a tailored environment to analize the nature of the EW phase transition via such a process, a study we 
leave for the future (see also~\cite{Tesietal} for a preliminary study in this direction).

\section{Conclusions}
\label{conclusionsNS}

Among the primary goals of future collider facilities is the precise analysis of the properties of the Higgs
sector. We have shown in this work that a high-energy $e^+ e^-$ machine like the proposed Compact Linear Collider -- CLIC -- operating at multi-TeV c.o.m. energies
would yield very sensitive direct probes of the existence of new scalars, combining the energy reach with the clean  
environment of an electron-positron machine. In particular, resonant di-Higgs searches in the $4 b$ final state at CLIC  
would surpass the reach of the HL-LHC by up to two orders of magnitude in the entire mass range 
$m_H \in [250\,\mathrm{GeV},\, 1 \,\mathrm{TeV}]$.
At the same time, these searches provide a direct avenue to probe the nature of the EW phase transition for non-minimal scalar sectors, 
and the possible origin of the cosmic matter-antimatter asymmetry via EW baryogenesis.

In the context of the extension of the SM by a real scalar singlet (SM + $S$, which could be viewed as a simple limit of the NMSSM or Twin Higgs theories), 
we have studied the sensitivity of CLIC to the parameter space where a strongly first order EW phase transition, as needed for successful baryogenesis, 
is realized. Our results show that there is a strong complementarity between direct searches for heavy Higgs bosons at CLIC via di-Higgs signatures, 
searches for heavy Higgses in di-boson ($WW$ and $ZZ$) final states at both HL-LHC and CLIC, and indirect probes of BSM physics via measurements 
of the Higgs self-coupling $\lambda_{111}$ and the Higgs associated production cross section $\sigma_{Zh}$
at CLIC and other future colliders like {\sl FCC-ee}. Combining the information from these searches could then allow to unravel the 
nature of EW symmetry breaking in the early Universe, and shed light on the origin of the baryon asymmetry of the Universe.

\acknowledgments

We would like to thank Roberto Franceschini and Andrea Tesi for useful discussions and comments, and Ulrike Schnoor for guidance on the CLIC Delphes implementation. 
We also thank Dario Buttazzo, Diego Redigolo, Filippo Sala and Andrea Tesi for correspondence regarding their forthcoming work~\cite{Tesietal}. 
J.M.N. is grateful to the Mainz Institute of Theoretical Physics (MITP) for its hospitality and 
its partial support during the completion of this work. 
J.M.N. was partially supported by the European Research Council under the European Union’s Horizon 2020 program, ERC
Grant Agreement 648680 (DARKHORIZONS) and by the Programa Atraccion de Talento de la Comunidad de Madrid under grant 2017-T1/TIC-5202, 
and also acknowledges support from the Spanish MINECO's ``Centro de Excelencia Severo Ochoa" Programme under grant SEV-2012-0249.

\bibliographystyle{JHEP.bst}

\begin{thebibliography}{10}


\bibitem{Khachatryan:2016vau}
  G.~Aad {\it et al.} [ATLAS and CMS Collaborations],
  JHEP {\bf 1608} (2016) 045
  [arXiv:1606.02266 [hep-ex]].
  
 \bibitem{ATLAS:2017ovn}
  The ATLAS collaboration [ATLAS Collaboration],
  ATLAS-CONF-2017-047.

   \bibitem{CMS:2018lkl}
  CMS Collaboration [CMS Collaboration],
  CMS-PAS-HIG-17-031.

  \bibitem{Morrissey:2012db}
  D.~E.~Morrissey and M.~J.~Ramsey-Musolf,
  New J.\ Phys.\  {\bf 14} (2012) 125003
  [arXiv:1206.2942 [hep-ph]].



\bibitem{Aicheler:2012bya}
  M.~Aicheler {\it et al.},
  doi:10.5170/CERN-2012-007

  \bibitem{CLIC:2016zwp}
  M.~J.~Boland {\it et al.} [CLIC and CLICdp Collaborations],
  doi:10.5170/CERN-2016-004
  arXiv:1608.07537 [physics.acc-ph]. 
  
  \bibitem{Profumo:2007wc}
  S.~Profumo, M.~J.~Ramsey-Musolf and G.~Shaughnessy,
  JHEP {\bf 0708} (2007) 010
  [arXiv:0705.2425 [hep-ph]]. 
  
 \bibitem{Barger:2007im}
  V.~Barger, P.~Langacker, M.~McCaskey, M.~J.~Ramsey-Musolf and G.~Shaughnessy,
  Phys.\ Rev.\ D {\bf 77} (2008) 035005
  [arXiv:0706.4311 [hep-ph]]. 
  
 \bibitem{Espinosa:2011ax}
  J.~R.~Espinosa, T.~Konstandin and F.~Riva,
  Nucl.\ Phys.\ B {\bf 854} (2012) 592
  [arXiv:1107.5441 [hep-ph]]. 
  
  
 \bibitem{Ellwanger:2009dp}
  U.~Ellwanger, C.~Hugonie and A.~M.~Teixeira,
  Phys.\ Rept.\  {\bf 496} (2010) 1
  [arXiv:0910.1785 [hep-ph]].
  
 \bibitem{Chacko:2005pe} 
  Z.~Chacko, H.~S.~Goh and R.~Harnik,
  Phys.\ Rev.\ Lett.\  {\bf 96}, 231802 (2006)
  [hep-ph/0506256]. 
  
  
  
  \bibitem{MoortgatPick:2005cw}
  G.~Moortgat-Pick {\it et al.},
  Phys.\ Rept.\  {\bf 460} (2008) 131
  [hep-ph/0507011].
  
  
 
  
  
 \bibitem{Alwall:2014hca}
  J.~Alwall {\it et al.},
  JHEP {\bf 1407} (2014) 079
  [arXiv:1405.0301 [hep-ph]].   
  
  
  \bibitem{Sjostrand:2014zea}
  T.~Sjöstrand {\it et al.},
  Comput.\ Phys.\ Commun.\  {\bf 191} (2015) 159
  [arXiv:1410.3012 [hep-ph]].
  
  \bibitem{deFavereau:2013fsa}
  J.~de Favereau {\it et al.} [DELPHES 3 Collaboration],
  JHEP {\bf 1402} (2014) 057
  [arXiv:1307.6346 [hep-ex]].
  
  \bibitem{UlrikeGitHub}
https://github.com/uschnoor/delphes.git
  
  \bibitem{AlipourTehrani:2254048}
N.~Alipour~Tehrani {\it et al.}, CLICdp-Note-2017-001 (March, 2017).
  
    \bibitem{Potter:2016pgp}
  C.~T.~Potter,
  arXiv:1602.07748 [hep-ph].
  
   
  
\bibitem{Aaboud:2017rel}
  M.~Aaboud {\it et al.} [ATLAS Collaboration],
  Eur.\ Phys.\ J.\ C {\bf 78} (2018) 293
  [arXiv:1712.06386 [hep-ex]].
  
 \bibitem{Heinemeyer:2013tqa}
  S.~Heinemeyer {\it et al.} [LHC Higgs Cross Section Working Group],
  doi:10.5170/CERN-2013-004
  arXiv:1307.1347 [hep-ph].
 
 
\bibitem{Aaboud:2017gsl}
  M.~Aaboud {\it et al.} [ATLAS Collaboration],
  Eur.\ Phys.\ J.\ C {\bf 78} (2018) no.1,  24
  [arXiv:1710.01123 [hep-ex]]. 
  

\bibitem{Dolan:2012rv} 
  M.~J.~Dolan, C.~Englert and M.~Spannowsky,
  JHEP {\bf 1210}, 112 (2012)
  [arXiv:1206.5001 [hep-ph]].  
  
\bibitem{deLima:2014dta} 
  D.~E.~Ferreira de Lima, A.~Papaefstathiou and M.~Spannowsky,
  JHEP {\bf 1408}, 030 (2014)
  [arXiv:1404.7139 [hep-ph]].  
  
  
\bibitem{Cacciari:2011ma}
  M.~Cacciari, G.~P.~Salam and G.~Soyez,
  Eur.\ Phys.\ J.\ C {\bf 72} (2012) 1896
  [arXiv:1111.6097 [hep-ph]].  
  
  
  \bibitem{Boronat:2014hva}
  M.~Boronat, J.~Fuster, I.~Garcia, E.~Ros and M.~Vos,
  Phys.\ Lett.\ B {\bf 750} (2015) 95
  [arXiv:1404.4294 [hep-ex]].
  
  \bibitem{CMS:2017xxp}
  CMS Collaboration [CMS Collaboration],
  CMS-PAS-HIG-17-009.
  

  
 \bibitem{No:2013wsa}
  J.~M.~No and M.~Ramsey-Musolf,
  Phys.\ Rev.\ D {\bf 89} (2014) no.9,  095031
  [arXiv:1310.6035 [hep-ph]]. 
  
  

  
\bibitem{Profumo:2014opa}
  S.~Profumo, M.~J.~Ramsey-Musolf, C.~L.~Wainwright and P.~Winslow,
  Phys.\ Rev.\ D {\bf 91} (2015) no.3,  035018
  [arXiv:1407.5342 [hep-ph]].  
  
\bibitem{Chen:2014ask}
  C.~Y.~Chen, S.~Dawson and I.~M.~Lewis,
  Phys.\ Rev.\ D {\bf 91} (2015) no.3,  035015
  [arXiv:1410.5488 [hep-ph]].  
  
 \bibitem{Robens:2015gla}
  T.~Robens and T.~Stefaniak,
  Eur.\ Phys.\ J.\ C {\bf 75} (2015) 104
  [arXiv:1501.02234 [hep-ph]].


  
\bibitem{Buttazzo:2015bka}
  D.~Buttazzo, F.~Sala and A.~Tesi,
  JHEP {\bf 1511} (2015) 158
  [arXiv:1505.05488 [hep-ph]].  
  
\bibitem{Bojarski:2015kra}
  F.~Bojarski, G.~Chalons, D.~Lopez-Val and T.~Robens,
  JHEP {\bf 1602} (2016) 147
  [arXiv:1511.08120 [hep-ph]].   
  
  
\bibitem{Kotwal:2016tex}
  A.~V.~Kotwal, M.~J.~Ramsey-Musolf, J.~M.~No and P.~Winslow,
  Phys.\ Rev.\ D {\bf 94} (2016) no.3,  035022
  [arXiv:1605.06123 [hep-ph]].  
  
  
\bibitem{Huang:2016cjm}
  P.~Huang, A.~J.~Long and L.~T.~Wang,
  Phys.\ Rev.\ D {\bf 94} (2016) no.7,  075008
  [arXiv:1608.06619 [hep-ph]].    
  
  
\bibitem{Huang:2017jws}
  T.~Huang, J.~M.~No, L.~Pernie, M.~Ramsey-Musolf, A.~Safonov, M.~Spannowsky and P.~Winslow,
  Phys.\ Rev.\ D {\bf 96} (2017) no.3,  035007
  [arXiv:1701.04442 [hep-ph]].  
  
  
\bibitem{Chen:2017qcz}
  C.~Y.~Chen, J.~Kozaczuk and I.~M.~Lewis,
  JHEP {\bf 1708} (2017) 096
  [arXiv:1704.05844 [hep-ph]].  
  
\bibitem{Lewis:2017dme}
  I.~M.~Lewis and M.~Sullivan,
  Phys.\ Rev.\ D {\bf 96} (2017) no.3,  035037
  [arXiv:1701.08774 [hep-ph]].

\bibitem{Curtin:2014jma}
  D.~Curtin, P.~Meade and C.~T.~Yu,
  JHEP {\bf 1411} (2014) 127
  [arXiv:1409.0005 [hep-ph]].  
  
\bibitem{Chala:2016ykx}
  M.~Chala, G.~Nardini and I.~Sobolev,
  Phys.\ Rev.\ D {\bf 94} (2016) no.5,  055006
  [arXiv:1605.08663 [hep-ph]].  
  
  
  
\bibitem{Dolan:1973qd}
  L.~Dolan and R.~Jackiw,
  Phys.\ Rev.\ D {\bf 9} (1974) 3320.
  
\bibitem{Nielsen:1975fs}
  N.~K.~Nielsen,
  Nucl.\ Phys.\ B {\bf 101} (1975) 173.  
  
\bibitem{Patel:2011th}
  H.~H.~Patel and M.~J.~Ramsey-Musolf,
  JHEP {\bf 1107} (2011) 029
  [arXiv:1101.4665 [hep-ph]].    
  
 
\bibitem{Wainwright:2011kj}
  C.~L.~Wainwright,
  Comput.\ Phys.\ Commun.\  {\bf 183} (2012) 2006
  [arXiv:1109.4189 [hep-ph]].  
  https://github.com/clwainwright/CosmoTransitions. http://clwainwright.github.io/CosmoTransitions.
  
\bibitem{ATL-PHYS-PUB-2014-016} 
 ATLAS-PHYS-PUB-2014-016
  
 \bibitem{Abramowicz:2016zbo}
  H.~Abramowicz {\it et al.},
  Eur.\ Phys.\ J.\ C {\bf 77} (2017) no.7,  475
  [arXiv:1608.07538 [hep-ex]]. 
  
\bibitem{LCD-Note-2012-014} 
T. Lastovicka and J. Strube, LCD-Note-2012-014  

\bibitem{Gomez-Ceballos:2013zzn}
  M.~Bicer {\it et al.} [TLEP Design Study Working Group],
  JHEP {\bf 1401} (2014) 164
  [arXiv:1308.6176 [hep-ex]].


\bibitem{dEnterria:2016fpc}
  D.~d'Enterria,
  Frascati Phys.\ Ser.\  {\bf 61} (2016) 17
  [arXiv:1601.06640 [hep-ex]].
  
  \bibitem{Craig:2013xia}
  N.~Craig, C.~Englert and M.~McCullough,
  Phys.\ Rev.\ Lett.\  {\bf 111} (2013) no.12,  121803
  [arXiv:1305.5251 [hep-ph]].
  
 \bibitem{Tesietal}
  D.~Buttazzo, D.~Redigolo, F.~Sala and A.~Tesi,
  to appear. 
  
  
\end{thebibliography}

\providecommand{\href}[2]{#2}\begingroup\raggedright\endgroup

 \end{document}